\documentclass[preprints,review,accept,pdftex,moreauthors]{Definitions/mdpi} 
\firstpage{1} 
\pubvolume{1}
\issuenum{1}
\articlenumber{0}
\pubyear{2022}
\copyrightyear{2022}
\datereceived{} 
\dateaccepted{} 
\datepublished{} 
\hreflink{https://doi.org/} % If needed use \linebreak

\usepackage[normalem]{ulem}

\Title{Recent Progress in Modelling the Macro- and Micro-Physics of Radio Jet Feedback in Galaxy Clusters}

\TitleCitation{Recent Progress in Modelling the Macro- and Micro-Physics of Radio Jet Feedback in Galaxy Clusters}

\Author{Martin A. Bourne $^{1, 2\ddagger}$*\orcidA{} and Hsiang-Yi Karen Yang $^{3, 4, 5,\ddagger}$*\orcidB{}}

\AuthorNames{Martin A. Bourne and Hsiang-Yi Karen Yang}

\AuthorCitation{Bourne, M. A.; Yang, H.-Y. K.}
\address{%
$^{1}$ \quad Institute of Astronomy, University of Cambridge, Madingley Road, Cambridge CB3 0HA, UK\\
$^{2}$ \quad Kavli Institute for Cosmology, University of Cambridge, Madingley Road, Cambridge CB3 0HA, UK\\
$^{3}$ \quad Institute of Astronomy and Department of Physics, National Tsing Hua University, Hsinchu 30013, Taiwan\\
$^{4}$ \quad Center for Informatics and Computation in Astronomy, National Tsing Hua University, Hsinchu 30013, Taiwan\\
$^{5}$ \quad Physics Division, National Center for Theoretical Sciences, Taipei 10617, Taiwan}

\corres{Correspondence: mabourne@ast.cam.ac.uk (M.A.B.), hyang@phys.nthu.edu.tw (H.-Y.K.Y)}

\secondnote{These authors contributed equally to this work.}
\abstract{Radio jets and the lobes they inflate are common in cool-core clusters and are known to play a critical role in regulating the heating and cooling of the intracluster medium (ICM). This is an inherently multi-scale problem, and much effort has been made to understand the processes governing the inflation of lobes and their impact on the cluster, as well as the impact of the environment on the jet-ICM interaction, on both macro- and microphysical scales. Developments of new numerical techniques and improving computational resources have seen simulations of jet feedback in galaxy clusters become ever more sophisticated. This ranges from modelling ICM plasma physics processes such as the effects of magnetic fields, cosmic rays and viscosity to including jet feedback in cosmologically evolved cluster environments in which the ICM thermal and dynamic properties are shaped by large-scale structure formation. In this review, we discuss the progress made over the last $\sim$~decade in capturing both the macro- and microphysical processes in numerical simulations, highlighting both the current state of the field as well as open questions and potential ways in which these questions can be addressed in the future.}


\begin{document}

\section{Overview}
\label{sec:overview}

\subsection{Jet Feedback in Galaxy Clusters: Observations and Theoretical Motivation}

Clusters of galaxies are the largest gravitationally collapsed objects in the Universe. Because of their deep gravitational potentials, the virial temperature of gas falling into clusters during hierarchical structure formation is heated to $10^{7-8}$ K, producing Bremsstrahlung emission in X-rays. Because the Bremsstrahlung emissivity of the intracluster medium (ICM) is proportional to gas density squared, near the centre of many massive clusters where the central density is high, the gas would cool significantly due to the strong radiation -- so-called ``radiative cooling.'' For clusters with central cooling times much smaller than the Hubble time, i.e., the cool-core (CC) clusters, it is predicted by the cooling-flow model \citep{Fabian1994} that there should be a significant inflow of gas toward the cluster centre, triggering intensive star formation within the central brightest cluster galaxies (BCGs). However, there is a lack of observational evidence for such cooling flows, and the star formation rates (SFRs) of the BCGs are typically 10-100 times lower than those predicted by the cooling-flow model. This is well-known as the ``cooling-flow problem'' of galaxy clusters \citep[see][for a review]{McNamaraNulsen2007, Fabian2012}.   

The absence of cooling flows calls for some heating mechanisms to balance radiative cooling in CC clusters, among which energetic feedback provided by relativistic jets emerging from the central supermassive black holes (SMBHs) is believed to be the most promising mechanism. This is motivated by the prevalence of X-ray cavities or bubbles, which are low-density, X-ray dim regions filled with radio-emitting plasma injected by the SMBH jets, in CC clusters \citep[e.g.,][]{BestEtAl2007, MittalEtAl2009, BirzanEtAl2012}, as well as in many galaxy groups \citep[e.g.,][]{BirzanEtAl2004, DavidEtAl2011, RandallEtAl2011, RandallEtAl2015}, which themselves represent a significant fraction of local radio galaxy hosts \citep[e.g.,][]{Best2004, CrostonEtAl2008, ChingEtAl2017}. Another compelling piece of evidence for active galactic nucleus (AGN) feedback is the observed correlation between the cavity power, which is a proxy for the jet power, and the X-ray luminosity within the cluster cores \citep{RaffertyEtAl2006, DunnFabian2008, HlavacekLarrondoEtAl2012, HlavacekLarrondoEtAl2022}, suggesting that there is a global thermal balance between heating and cooling in CC clusters. These observations have therefore motivated a picture of a self-regulated AGN feedback loop: when a cluster has increased X-ray luminosity, the ICM cooling rates and thus SMBH accretion rates are enhanced, which triggers powerful ejections of relativistic jets, heating the ICM and shutting off subsequent cooling and AGN activity \citep{McNamaraNulsen2012}. While the idea of an AGN feedback cycle is an attractive solution to the cooling-flow problem, there is a huge dynamical range between the scales of SMBH accretion discs ($\sim 1-10^4$ AU) and the cores of galaxy clusters ($\sim 100$ kpc) (see Fig.\ 1 in \citet{GaspariEtAl2020} for an illustration of the multi-scale SMBH accretion and feedback processes involved). The vast dynamical range involved has prohibited an all-in-one simulation of AGN feedback, and hence the detailed processes of SMBH feeding and feedback mechanisms remain one of the greatest unsolved problems in astrophysics.

Because galaxy clusters are at the crossroads of cosmology and astrophysics, the advantages of understanding AGN feedback in clusters are two-fold. On the one hand, galaxy clusters are excellent laboratories for studying various astrophysical phenomena, including radiative cooling, AGN feedback, magnetic fields, shocks, turbulence, acceleration of cosmic rays (CRs), plasma physics, etc. On the other hand, constraining these astrophysical mechanisms would enable robust predictions of cluster observables and evolution histories that are critical for inferences of cosmological parameters as well as the formation and evolution of galaxies \citep[see][for relevant reviews]{KravtsovBorgani2012, AllenEtAl2011}. In turn, the large-scale environments of clusters and their evolution histories would also influence the operation of the above astrophysical processes. AGN jet feedback in clusters is one great example of this iterative process, where the theoretical understanding from both the astrophysical and cosmological perspectives has emerged in the past two decades. It is therefore our aim to review this exciting progress in this article. AGN feedback is not limited to jets in clusters, additionally manifesting itself via radiation and accretion disc driven winds \citep{Fabian2012, KingPounds2015, Morganti2017}, and is expected to impact galaxy formation over a wide range of redshifts by shaping galaxy properties, regulating BH growth and driving BH-host galaxy scaling relations \citep[][]{MagorrianEtAl1998, FerrareseMerritt2000, KormendyHo2013, McConnellMa2013, SahuEtAl2019}, which themselves may aid our understanding of AGN physics. AGN feedback may even be important at the low mass end, with recent studies of its affect on dwarf galaxies representing an area of growing interest \citep[e.g.,][]{ReinesEtAl2013, Manzano-KingEtAl2019, Mezcua2021, KoudmaniEtAl2022, SharmaEtAl2022}. We refer interested readers to additional reviews that provide comprehensive coverage of these topics \citep[e.g.][]{McNamaraNulsen2007, Fabian2012, McNamaraNulsen2012, KormendyHo2013, KingPounds2015, Harrison2017, Morganti2017, HarrisonEtAl2018, HardcastleCroston2020, VeilleuxEtAl2020, Mezcua2021, HlavacekLarrondoEtAl2022}.

\subsection{Simulating Jets in Idealised Hydrodynamic Simulations}
\label{sec:sim_overview}

As early as the late 1970s and throughout the 1980s, groups were performing simulations of jets to study their propagation and structural evolution (see e.g.,\citet{BurnsEtAl1991} for a review at the time). The pioneering works of \citet{Rayburn1977} and \citet{NormanEtAl1982} paved the way, performing the first simulations of non-relativistic jets. Numerous works followed with a particular emphasis on understanding radio observations of jets \citep{ClarkeEtAl1989, ClarkeEtAl1992, FalleWilson1985, SmithEtAl1985, WilsonScheuer1983}, as well as numerical studies on the impact of resolution \citep{KoesslMueller1988}, that included magnetic fields \citep{ClarkeEtAl1986, ClarkeEtAl1989, LindEtAl1989} and that were performed in 3D \citep{ArnoldArnett1986, ClarkeEtAl1990}. The history of simulating jets is itself an interesting topic, however, in this work, we restrict ourselves to more recent advances in modelling jet feedback in galaxy cluster environments and instead point the reader to reviews focusing on simulating jets in general \citep{KomissarovPorth2021, Marti2019}.

While much early work was motivated by radio observations of jets \citep[e.g.,][]{Miley1980, BirettaEtAl1983}, X-ray observations of galaxy clusters using ROSAT \citep[e.g.,][]{BoehringerEtAl1993, CarilliEtAl1994, HuangEtAl1998} and later Chandra \citep[e.g.,][]{FabianEtAl2000, McNamaraEtAl2000} that show cavities\footnote{As highlighted by \citet{Fabian2012}, cavities had been observed earlier with the Einstein telescope \citep{BranduardiRaymontEtAl1981, FabianEtAl1981}, although their true nature had not been realised.} in the ICM provided further motivation to understand the interaction of radio jets and the lobes they inflate with the surrounding hot ICM. Broadly speaking, two approaches were used to study this scenario; one in which the jets were simulated and the self-consistent inflation of the lobes studied \citep{ClarkeEtAl1997, RizzaEtAl2000, ReynoldsEtAl2001, ReynoldsEtAl02, OmmaEtAl2004}, or for simplicity, one in which this process was pre-assumed and hot bubbles were added or inflated ``by hand'' \citep{ChurazovEtAl2001, BruggenKaiser2001, QuilisEtAl2001, BruggenEtAl2002}. While the former approach gives insights into how the jet itself interacts with the ICM and how lobes are inflated, the latter is less computationally expensive. More complex simulations have followed and contributed to an ever-growing literature, which we discuss further throughout this review.

Apart from a few examples \citep{HuskoLacey2023Method, HuskoLacey2023Interplay, HuskoEtAl2022}, simulations that attempt to model the jets themselves are performed using grid-based codes that allow for high resolution in low-density regions, which are typical of jets. Traditionally such codes are Eulerian, although novel refinement techniques have meant that moving-mesh \citep[AREPO, ][]{Springel2010Arepo} and meshless-finite-mass \citep[GIZMO, ][]{Hopkins2015Gizmo} codes have also been used to simulate jets \citep[e.g.,][]{BourneSijacki2017, WeinbergerEtAl2017, SuEtAl2021}. When it comes to launching a jet, some works place finer control on how exactly the jet is injected into the simulation domain, by defining the exact thermodynamic and kinetic state within the launch region, i.e., the jet density, temperature and velocity \citep[e.g.,][]{Krause2003, KrauseEtAl2012, HardcastleKrause2013, WeinbergerEtAl2017}. Other works simply add mass, momentum and/or energy to the injection region and allow the jets' thermodynamic properties to arise naturally from this \citep[e.g.][]{OmmaEtAl2004, CattaneoTeysier2007, DuboisEtAl2010, GaspariEtAl2011a, LiBryan2014, YangReynolds2016Hydro, BourneSijacki2017}. However, the ultimate result is a pair of fast low density jets that drive bow shocks into the ICM (and potentially internal shocks along the jets), thermalise, and inflate high-temperature lobes that come into pressure balance with the ICM. These methods have been used to perform a wide range of simulations, from high-resolution studies of lobe inflation and energetics \citep[e.g.,][]{HardcastleKrause2013, HardcastleKrause2014, BourneSijacki2017, WeinbergerEtAl2017, EhlertEtAl2018, BourneEtAl2019, YangEtAl2019, PeruchoEtAl2022} to those that include SMBH accretion models from which the jet power is calculated \citep[e.g.,][]{GaspariEtAl2011a, GaspariEtAl2011b, GaspariEtAl2015, PrasadEtal2015, YangReynolds2016Hydro, YangReynolds2016Conduction, LiEtAl2017, BeckmannEtAl2019, EhlertEtAl2022} and are performed on Gyr time-scales to understand how jets can regulate the thermodynamic state of the ICM. On top of this, additional physics relative to jet feedback and galaxy clusters has been included in dedicated jet simulations such as magnetic fields \citep{RobinsonEtAl2004, LiEtAl2006, RuszkowskiEtAl2008, DuboisEtAl2009, GaiblerEtAl2009, OneillJones2010, MendygralEtAl2012, SutterEtAl2012, HardcastleKrause2014, WeinbergerEtAl2017, EhlertEtAl2018, WangEtAl2020, WangEtAl2021, EhlertEtAl2022}, relativistic effects \citep{PeruchoEtAl2014, EnglishEtAl2016, PeruchoEtAl2017, EnglishEtAl2019, YatesEtAl2021, PeruchoEtAl2022}, viscosity \citep{ReynoldsEtAl2005, DongStone2009, Guo2015, KingslandEtAl2019}, thermal conduction \citep{RuszkowskiBegelman2002, ZakamskaNarayan2003, VoigtFabian2004, ParrishEtAl2009, BogdanovicEtAl2009, AvaraEtAl2013, YangReynolds2016Conduction, KannanEtAl2017, BeckmannEtAl2022Conduction} and CR physics \citep{GuoOh2008, MathewsBrighenti2008, RuszkowskiEtAl2017b, WeinbergerEtAl2017, BourneSijacki2017, JacobPfrommer2017a, JacobPfrommer2017b, ProkhorovChurazov2017, EhlertEtAl2018, YangEtAl2019, LinEtAl2023}.

There are now many works that have coupled jet power to SMBH accretion rates \citep[e.g.,][]{CattaneoTeysier2007, DuboisEtAl2010, DuboisEtAl2012, GaspariEtAl2011a, GaspariEtAl2011b, GaspariEtAl2012, YangEtAl2012, LiBryan2014, LiEtAl2017, YangReynolds2016Hydro, PrasadEtal2015, MeeceEtAl2017, EhlertEtAl2022}. The physical connection between accretion and jet power, expected to depend on the properties of the BH and its accretion flow \citep[see e.g. Section 4 of][for a discussion]{HlavacekLarrondoEtAl2022}, is still to be fully understood. For simplicity, the simulations discussed here often assume a fixed scaling between accretion rate, $\dot{M}$, and jet power, $\dot{E}_{\rm Jet}$, of the form 
\begin{equation}
\dot{E}_{\rm Jet}=\epsilon\dot{M}c^{2},
\label{eq:mdot}
\end{equation}
where $c$ is the speed of light and $\epsilon$ is a constant feedback efficiency, although variable spin-dependent efficiencies have been included in recently developed models \citep{BeckmannEtAl2019, TalbotEtAl2021, TalbotEtAl2022, HuskoEtAl2022}. \citet{CattaneoTeysier2007} presented some of the earliest simulations that coupled a jet feedback scheme \citep{OmmaEtAl2004} to a Bondi accretion model and therefore capture self-regulated jet feedback in a galaxy cluster. The jets were injected using a combination of momentum and thermal energy and while the feedback was able to prevent over-cooling of the ICM it was overly effective and unable to preserve a CC. Numerous works followed that similarly investigated injecting a combination of thermal energy and momentum \citep{YangEtAl2012, LiBryan2014} as well as works that considered purely kinetic jets \citep{GaspariEtAl2011b, GaspariEtAl2012, PrasadEtal2015, YangReynolds2016Hydro, YangReynolds2016Conduction} that were able to inhibit cooling and match a range of observed cluster properties. Specifically, it has been found that combining momentum-driven jets with cold-accretion models \citep{GaspariEtAl2012, GaspariEtAl2013, GaspariEtAl2015, LiBryan2014, LiEtAl2017, YangReynolds2016Hydro, PrasadEtal2015, MeeceEtAl2017, EhlertEtAl2022}, whereby the SMBH grows due to accretion of cold gas, can regulate ICM cooling while maintaining the properties of CC clusters, albeit in idealised cluster environments. However, as discussed in the next section, reproducing thermodynamic profiles and the CC/non-CC dichotomy is difficult to achieve in full cosmological simulations.

\subsection{AGN Feedback in Cosmological Simulations}
\label{sec:CosmoSims}

The simulations discussed above, typically performed in hydrostatic atmospheres, place emphasis on understanding the detailed processes governing lobe inflation and interaction with the ICM, which requires high spatial (often achieved with dedicated lobe refinement techniques) and time resolution. However, the study of cluster formation and evolution requires simulations that capture structure formation processes over cosmic time, i.e. cosmological simulations. Such simulations span a wide range of galaxy masses and include various models to capture a wide array of physical processes important to galaxy formation and we point the reader to reviews for a detailed discussion \citep{SomervilleDave2015, VogelsbergerEtAl2020}. The inclusion of AGN feedback in cosmological simulations has been shown to be an important ingredient to create realistic galaxy populations, particularly at the high mass end \citep[e.g.,][]{DiMatteoEtAl2005, LeBrunEtAl2014, SchayeEtAl2015, SijackiEtAl2015, DuboisEtAl2016, WeinbergerEtAl2018}. There are a growing number of cosmological simulations that capture the formation and evolution of galaxy clusters and aim to understand the processes that shape their properties \citep[e.g.,][]{DuboisEtAl2011, LeBrunEtAl2014, BaheEtAl2017, BarnesEtAl2017a, BarnesEtAl2017b, BarnesEtAl2018, HahnEtAl2017, RasiaEtAl2015, HendenEtAl2018, TremmelEtAl2019, PakmorEtAl2022}. Such simulations have done remarkably well at matching a range of observed cluster properties, however, achieving realistic thermodynamic profiles in cluster cores and explaining the CC/non-CC dichotomy is challenging, with several works concluding a need for modified AGN feedback models \citep[e.g.,][]{GenelEtAl2014, BarnesEtAl2017b, HendenEtAl2018}.

Cosmological simulations, by the necessity of space and time resolution constraints, implement simplified and/or low-resolution models of AGN feedback that do not necessarily capture the processes of lobe inflation and their interaction with the ICM in detail. In the most straightforward implementation, feedback energy is injected thermally in resolution elements local to the BH \citep[e.g.][]{SpringelEtAl2005, DiMatteoEtAl2005, BoothEtAl2009, SchayeEtAl2015, McCarthyEtAl2017, TremmelEtAl2019}. Modifications are often made to such models to circumvent the over-cooling problem \citep[see e.g.,][for a discussion]{BoothEtAl2009, LeBrunEtAl2014, BourneEtAl2015, SchayeEtAl2015}, for example by temporarily switching off radiative cooling \citep{TremmelEtAl2017, TremmelEtAl2019}, or by storing energy until gas can be heated to a minimum temperature \citep{BoothEtAl2009, LeBrunEtAl2014, SchayeEtAl2015, McCarthyEtAl2017} or a fixed duty cycle has passed \citep{HendenEtAl2018}. Alternatively, the energy can be injected as a kinetic outflow \citep{DuboisEtAl2010, DuboisEtAl2012, DuboisEtAl2016, WeinbergerEtAl2017Wind, WeinbergerEtAl2018, DaveEtAl2019}. While some works assume a single feedback mode that is the same irrelevant of the BH mass or accretion rate \citep[e.g.,][]{SchayeEtAl2015, McCarthyEtAl2017, TremmelEtAl2019}, others implement separate ``quasar'' and ``radio'' modes \citep[e.g.,][]{SijackiEtAl2007, SijackiEtAl2015, DuboisEtAl2012, DuboisEtAl2016, WeinbergerEtAl2017Wind, WeinbergerEtAl2018, DaveEtAl2019}, with the latter typically being used for low accretion rates. One of the earliest such examples of dual AGN modes was introduced by \citet{SijackiEtAl2007}, taking inspiration from early bubble models of jet feedback \citep{ChurazovEtAl2001, QuilisEtAl2001, DallaVechiaEtAl2004}, by including the jet mode as thermal bubbles displaced from the central SMBH. This model has subsequently been used in the Illustris \citep{SijackiEtAl2015} and Fable \citep{HendenEtAl2018} simulation suites. On the other hand, \citet{DuboisEtAl2010}, built on the work of \citet{OmmaEtAl2004} and \citet{CattaneoTeysier2007} to implement mass-loaded bipolar outflows that aimed to capture a sub-relativistic accretion disc wind as jet-like outflows and can self-consistently produce cavity like structures in the ICM. This model has since been used in the HorizonAGN \citep{DuboisEtAl2016} and NewHorizon \citep{DuboisEtAl2021} simulations, as well as other studies of feedback in galaxy clusters \citep[e.g.,][]{BeckmannEtAl2019, BeckmannEtAl2022Conduction}. In a somewhat similar vein, \citet{DaveEtAl2019} implement a jet feedback model for the low accretion rate regime in the form of a slow, highly collimated bipolar outflow, in the SIMBA suite of simulations, however, the jet is hydrodynamically decoupled during its initial phase, meaning that it does not interact with the inner parts of galaxies and clusters and instead couples its energy on larger scales. Alternatively, some simulation suites, such as IllustrisTNG \citep{WeinbergerEtAl2018}, take a different approach and instead of attempting to model jet feedback implement an efficient kinetic wind model for high mass SMBHs in the low accretion rate regime that injects kinetic energy into the local gas along a random direction \citep{WeinbergerEtAl2017Wind}, which acts to effectively quench high mass galaxies.

\subsection{The Importance of Macrophysics and Microphysics}
\label{sec:intro_importance}

Thanks to the advancement in computing power and numerical algorithms over the past decade, numerical simulations of jet feedback in clusters have become more and more sophisticated with realistic initial conditions from cosmological contexts as well as with complex physical processes including magnetic fields, CRs, plasma effects, etc. The latter is often dubbed as ``microphysics'' in clusters because the gyro-radii of charged particles and CRs in the magnetised ICM (on the order of $\sim$ AU for GeV CRs in $\mu$G magnetic fields) are many orders of magnitude smaller than the size of clusters ($\sim$ Mpc). Consequently, transport processes in the ICM, including thermal conduction, viscosity, and CR propagation, are determined by plasma physics that happen on the microscopic scales of the particle gyro-radii (Section~\ref{sec:micro}). As will be discussed in detail in later sections, recent simulations have demonstrated that both the macrophysics from the large-scale environments of clusters and the microphysics have crucial impacts on the AGN feeding and feedback processes and the evolution of clusters, and thus they will be the focus of this review. We also point out, however, that despite the vast computational capabilities now available to the community, there is still a need for robust, physically motivated analytic components and sub-grid models to link between the scales above and below the resolution limit of simulations.

The structure of this article is as follows. In Section~\ref{sec:macro}, we will review the important progress regarding the modelling of the macrophysical processes of AGN feedback. We will first start by introducing the fundamental processes of AGN lobe formation and lobe-ICM interaction (Section~\ref{sec:bubbles_shocks_waves}). This is followed by discussions about the role of the cluster environment (Section~\ref{sec:macro_environment}) and open questions and future opportunities (Section~\ref{sec:macro_future}). In Section~\ref{sec:micro}, important findings regarding the microphysical processes will be summarized. Specifically, in Section~\ref{sec:cr} we will discuss the roles of CRs in AGN feedback and their observational signatures. We will then discuss the roles of plasma physics, such as transport processes like conduction and viscosity in the ICM, in Section~\ref{sec:plasma}. Open questions and future opportunities regarding the microphysics are discussed in Section~\ref{sec:micro_future}. Finally, Section~\ref{sec:conclusions} contains our concluding remarks. 

\section{Modelling the Macrophysics}
\label{sec:macro}

\begin{figure}[H]
\includegraphics[width=13.5cm]{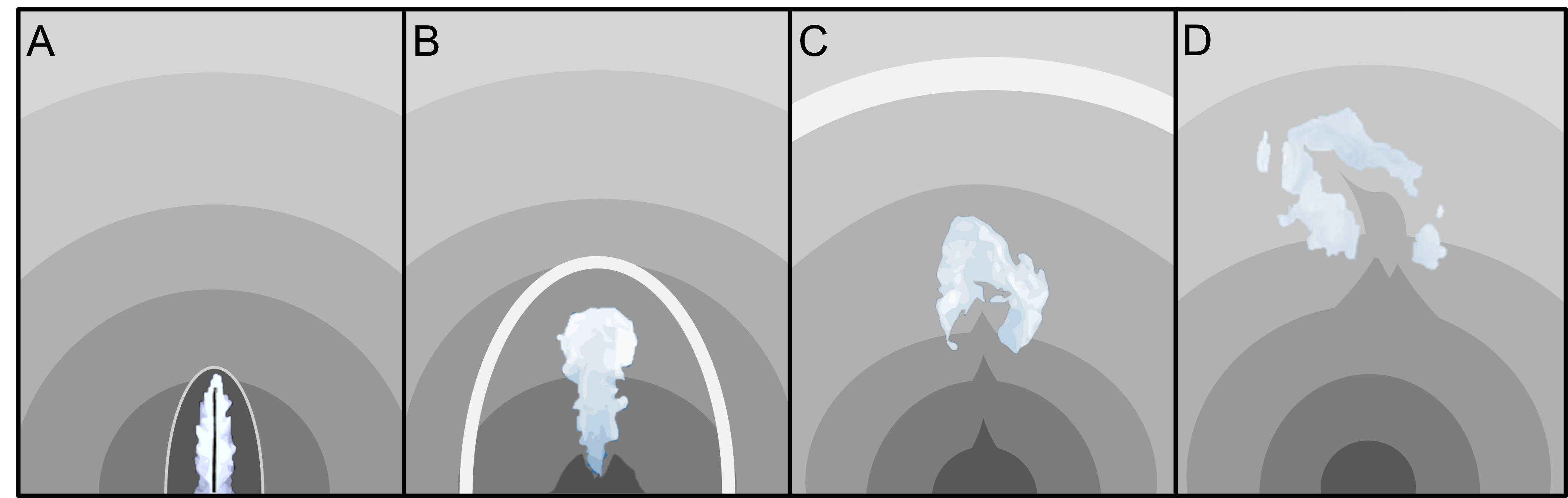}
\caption{Diagram illustrating the general processes that occur during and after lobe inflation. {\it Panel A:} A fast jet drives into the ambient medium, forms a bow shock and inflates a hot lobe that expands into the ICM. The lobe morphology can depend sensitively on the injected jet properties (e.g., content, velocity, geometry). The expanding shock wave results in a layer of shocked ICM material surrounding the jet lobe. {\it Panel B:} As the lobe expansion slows, which may or may not be accompanied by the jet switching off, the shock driven into the ICM broadens into a sound wave that can detach from the lobes. {\it Panel C:} Once the jet has ceased and as the lobe buoyantly rises through the ICM, dense, low entropy material can be entrained and pulled up in the wake and instabilities can lead to mixing of the lobe and ICM material. Sound waves generated by the lobe expansion can continue to propagate to large distances depending on the ICM viscosity. {\it Panel D:} This process continues at late times, with mixing continuing to dilute the lobe material, although the rate at which this occurs can depend sensitively on the ICM physical processes including magnetic fields, viscosity and cluster weather (i.e. ICM and sub-structure motions, see Section~\ref{sec:macro_environment})}
\label{fig:overview}
\end{figure}

There is now significant literature on hydrodynamic and magnetohydrodynamic (MHD) modelling of the large-scale processes driving the inflation of jet lobes and their macroscopic evolution. Figure~\ref{fig:overview} provides a general diagrammatic overview of the phases of lobe inflation and subsequent evolution, which we discuss throughout this section. Specifically, in Section~\ref{sec:bubbles_shocks_waves} we first consider the processes that are intrinsic to the jets and lobes, including their energetics and processes through which they interact with the ICM, while in Section~\ref{sec:macro_environment} we additionally discuss the role of the cluster environment, including cluster weather, and the role this has on shaping lobe properties and cluster heating.

\subsection{Bubbles, Shocks and Waves}
\label{sec:bubbles_shocks_waves}

\subsubsection{Morphology, Direction and Energetics}
\label{sec:morphology_direction_energetics}

\begin{figure}
\includegraphics[width=13.5cm]{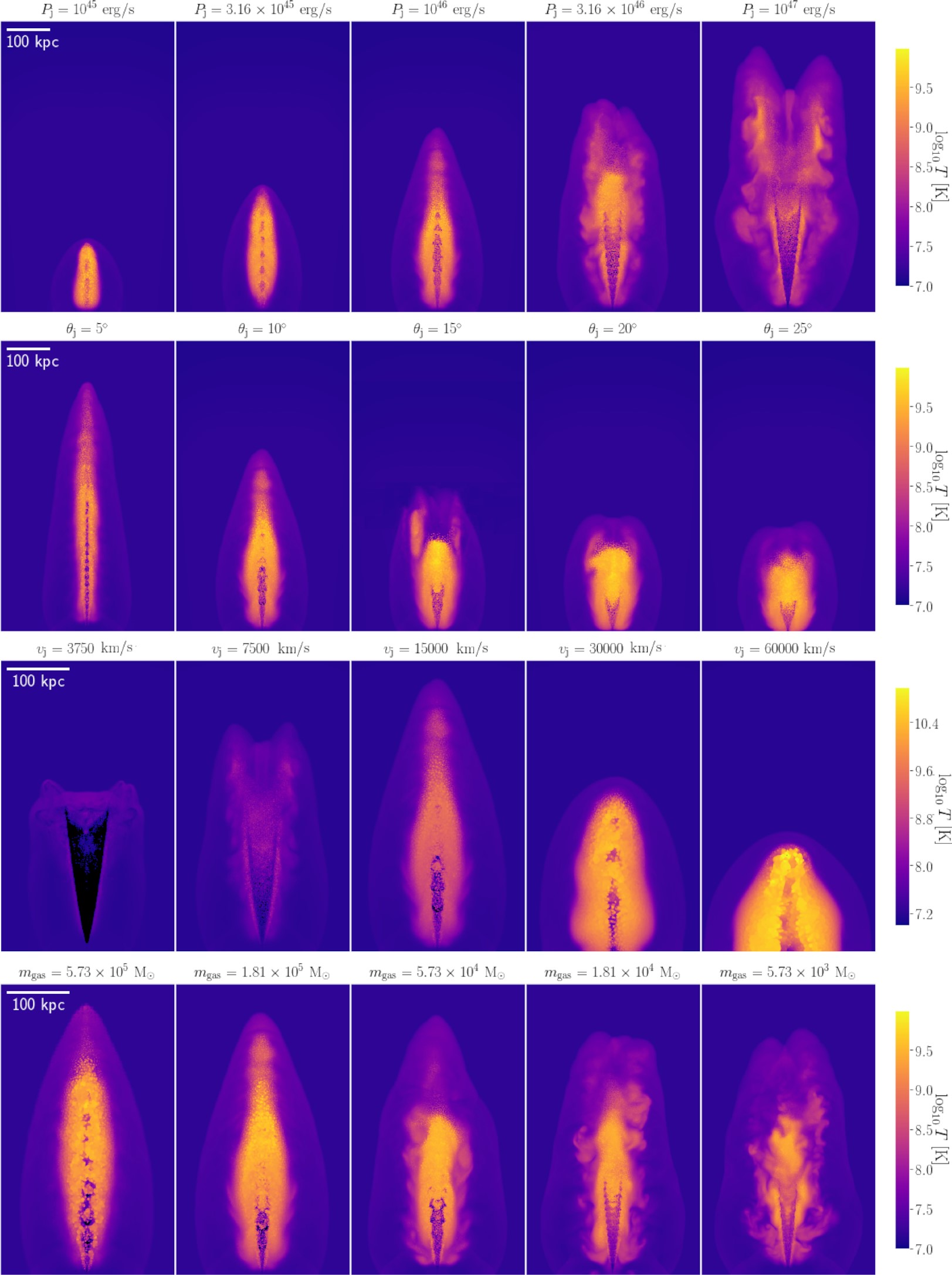}
\caption{Thin temperature projections illustrate how jet injection parameters impact jet and lobe morphologies. All jets are $100$~Myr old with each row illustrating the effect of changing one parameter, which from top to bottom are jet power, half opening angle, velocity and resolution, respectively. These quantities take fiducial values of $10^{46}$~erg s$^{-1}$, $10^{\circ}$, $15000$~km s$^{-1}$ and $1.81\times10^5$~M$_{\odot}$, unless being varied. Colour bars shown to the right of each row extend down to $10^7$~K, gas below this is shown in black. (Figure 6 from \citet{HuskoLacey2023Method}, CC BY 4.0)}
\label{fig:husko_injection}
\end{figure} 

Observed jets have long been classified based on their radio morphology as set out by \citet[FR][]{FanaroffRiley1974}, which broadly characterises them as centre-brightened FR-I sources \citep[e.g. 3C 449,][]{PerleyEtAl1979} that are expected to form due to decelerating and becoming turbulent on small ($\sim$~kpc) scales\footnote{Frustrated and bent jets, such as wide/narrow angle tailed radio galaxies, can provide morphological analogues to FR-I sources \citep[e.g.,][]{DehghanEtAl2014, OBrienEtAl2018, MurthyEtAl2022, MukherjeeEtAl2018}.} \citep[e.g.,][]{Komissarov1994, RossiEtAl2008, PeruchoEtAl2010, LaingBridle2014, TchekhovskoyBromberg2016}, or edge-brightened FR-II sources \citep[e.g. Cygnus A,][]{CarilliBarthel1996} that are highly collimated and expected to be relativistic along their entire length. Although the differences are expected to result from the dynamics of the jets and depend on both their power and how they interact with their environment, some debate remains over the exact role of the central engine \citep[see the review of][for a fuller discussion]{HardcastleCroston2020}, with a number of works, both theoretical and observational, considering the relationships between AGN properties, such as radio loudness, morphology and/or excitation, and the properties of the accretion flow and/or BH \citep{WilsonColbert1995, SikoraEtAl2007, TchekhovskoyEtAl2010, TchekhovskoyEtAl2012, BestHeckman2012, GendreEtAl2013, MingoEtAl2014, InesonEtAl2015, MingoEtAl2022}. Several parameter studies that vary the jet (e.g., power, opening angle, velocity) and environmental properties have been performed that can recover both FR-I and FR-II sources \citep[e.g.,][]{KrauseEtAl2012, TchekhovskoyBromberg2016, MassagliaEtAl2016, MassagliaEtAl2019, MassagliaEtAl2022, EhlertEtAl2018, LiEtAl2018, MandalEtAl2022, YatesEtAl2023}, with \citet{MassagliaEtAl2016} suggesting that simulating FR-I sources is challenging and requires high-resolution 3D simulations to fully capture their turbulent properties. 

As discussed in Section~\ref{sec:sim_overview}, several methods have been developed to launch jets and capture the lobe inflation process within the framework of galaxy cluster evolution. It was found early on that to create a cocoon structure the jet needs to be lower density than the surrounding medium \citep[see discussion in e.g.,][and references therein]{Ferrari1998, Krause2003, WeinbergerEtAl2017}. As outlined in Section~\ref{sec:sim_overview}, many studies now make use of fast collimated outflows that efficiently thermalise through shocks to inflate lobes of hot gas, however, a range of morphologies are still seen in simulations depending on both numerical and physical choices for how the jet is injected, as recently illustrated by \citet{HuskoLacey2023Method} and shown in Figure~\ref{fig:husko_injection}. The choice of using a cylindrical ($0^{\circ}$ opening angle) versus conical injection, as well as the jet opening angle itself, can impact jet properties \citep[e.g.,][]{Falle1991, KrauseEtAl2012, HuskoLacey2023Method, YatesEtAl2023} and is a possible determinant of whether the jet has FR-I or FR-II morphology \citep{KrauseEtAl2012, YatesEtAl2023}. Jets with a high momentum density, which occurs in slow/heavy jets\footnote{One can achieve the same jet power by either having fast, light jets or slow, heavy jets, with the latter having larger momenta than the former.} are typically elongated along the jet direction with narrow lobes, whereas light/fast jets are found to be shorter and inflate wide lobes that have a notable expansion perpendicular to the jet direction \citep[e.g.,][]{Krause2003, Krause2005, GaiblerEtAl2009, Guo2015, Guo2016, WeinbergerEtAl2017, SuEtAl2021, HuskoLacey2023Method, WeinbergerEtAl2022, YatesEtAl2023}. Although additionally, \citet{HuskoLacey2023Method} found that these trends can break down for very slow outflows that are unable to inflate hot lobes nor generate significant backflows (see third row of Figure~\ref{fig:husko_injection}). In terms of regulating cluster cooling and heating, the recent study of \citet{WeinbergerEtAl2022} highlighted that the behaviour of very light jets inflating wider lobes makes them more effective than heavy jets at removing low entropy material from cluster cores and inhibiting cooling flows \citep[see also,][]{SuEtAl2021}.

Some works have also considered if and how injecting magnetic fields with the jets can influence lobe morphology and dynamics \citep[e.g.,][]{GaiblerEtAl2009, HardcastleKrause2014, WeinbergerEtAl2017, EhlertEtAl2018, MassagliaEtAl2019, EnglishEtAl2016}. \citet{GaiblerEtAl2009} found that injected fields, amplified in the jet head, can stabilise against Kelvin-Helmholtz instabilities and generate a cleaner lobe structure. Works that consider magnetic fields with moderate to high values of plasma $\beta$ (defined as the ratio between thermal and magnetic pressure, $\beta \equiv P_{\rm th}/P_{\rm B}$) find, as expected, that they have limited impact on the overall lobe dynamics \citep{HardcastleKrause2014, WeinbergerEtAl2017, EhlertEtAl2018, EnglishEtAl2016} although they can affect the long-term stability of lobes by suppressing instabilities that would otherwise mix the jet and ICM material \citep[e.g.,][]{WeinbergerEtAl2017, EhlertEtAl2018}. On the other hand, \citet{MassagliaEtAl2019} show that strong fields ($\beta=3$) can impact morphology, leading to the jet becoming distorted due to non-axisymmetric modes. In terms of observational constraints, lobe magnetic fields have been measured using inverse-Compton observations of FR-II sources \citep[e.g.,][]{KataokaStawarz2005, CrostonEtAl2005, InesonEtAl2017, HarwoodEtAl2017}, with \citet{InesonEtAl2017} finding observed field strengths to be below equipartition in all of their sources with a median ratio of $0.4$ and hence very low values of $\beta$ would seem unlikely, although \citet{CrostonEtAl2005} do find some magnetically dominated sources. Another important factor that many simulations of jet feedback on galaxy cluster scales neglect is the effect of relativity, both in terms of the jet dynamics and in correctly accounting for high-temperature gas. For jet speeds approaching the speed of light, the effects of relativity are expected to play an important role in shaping lobe properties and increase the efficiency and volume over which jets can heat the ICM \citep{PeruchoEtAl2014, PeruchoEtAl2017, PeruchoEtAl2022}. Additionally, the behaviour that fast-light jets inflate wider lobes that interact more isotropically with the ICM is found to extend into the relativistic regime \citep{EnglishEtAl2016, YatesEtAl2023}, while simulations that consider CR-dominated jets also produce ``fat'' lobes (see Section~\ref{sec:cr}).

The mechanisms that launch a jet and determine its direction occur on scales of the accretion disc or smaller\footnote{A binary companion can also result in jet precession \citep{BegelmanEtAl1980, KrauseEtAl2019}, although the mechanisms governing this are also not resolved in large-scale cluster simulations.} \citep[e.g.,][]{TchekhovskoyEtAl2010, TchekhovskoyEtAl2011, SadowskiEtAl2015, LiskaEtAl2018, LiskaEtAl2019, LuEtAl2023}, which is far below the resolution limit of galaxy cluster simulations. As such, ab-initio modelling of the jet direction evolution is not possible. Traditionally, simulations of jets in galaxy clusters assume a fixed jet direction potentially with some small angle precession \citep[e.g.][]{Falceta-GoncalvesEtAl2010, GaspariEtAl2012, LiBryan2014, YangReynolds2016Hydro, BourneSijacki2017, MartizziEtAl2019}, although some works alternatively randomly re-orientate the jet direction by hand \citep[e.g.,][]{CieloEtAl2018}, to improve coupling between the jet and ICM. Several subgrid models have been developed recently, which while differing in the exact assumptions made regarding the subgrid accretion disc properties, can track BH spin evolution \citep[e.g.,][]{DuboisEtAl2014, FiacconiEtAl2018, BustmanteSpringel2019, BeckmannEtAl2022Conduction, HuskoEtAl2022}. \citet{DuboisEtAl2014} additionally used the SMBH spin to determine the radiative efficiencies and jet direction, while \citet{BeckmannEtAl2019} further developed this model to use spin-dependent jet efficiencies derived from GRMHD simulations \citep{McKinneyEtAl2012}. In a similar vein, \citet{TalbotEtAl2021} coupled a jet feedback model \citep{BourneSijacki2017} to the accretion disc model of \citet{FiacconiEtAl2018}, assuming spin-dependent Blandford-Znajek jet efficiencies also derived from GRMHD simulations \citep{TchekhovskoyEtAl2010} as well as the back reaction of the jet on the SMBH evolution. While the models of \citet{DuboisEtAl2014} and \citet{FiacconiEtAl2018} both assume a thin $\alpha$-disc, \citet{HuskoEtAl2022} assume a thick disc, which typically leads to slower spin alignment timescales. While such methods are not necessary for modelling single feedback events, simulations that capture self-regulated feedback on Gyr timescales and include spin evolution find that the jets can re-orient and inject energy more isotropically within the cluster \citep{BeckmannEtAl2019, HuskoEtAl2022}.

The energetics governing lobe inflation has been an important area of study in particular due to its consequence on how the feedback heats the ICM. Observational measurements of the lobe pressure and volume are often used to constrain lobe and jet energetics which show a correlation between the estimated lobe power and ICM cooling rate \citep{Fabian2012, HlavacekLarrondoEtAl2012, HlavacekLarrondoEtAl2022}. These lobe powers have also been compared to estimated \citet{Bondi1952} accretion rates to constrain the efficiency, i.e. $\epsilon$ in Equation~\ref{eq:mdot}, of the central engine \citep{AllenEtAl2006, MerloniHeinz2007, McNamaraEtAl2011, RussellEtAl2013, NemmenEtAl2015}, with some studies requiring $\epsilon>1$. Setting aside the uncertainties in estimating lobe powers and Bondi accretion rates from observations \citep[e.g.][]{RussellEtAl2013}, high efficiencies could indicate that modes other than Bondi are needed to feed the BH, such as cold accretion \citep[see e.g.,][for further discussion, including limitations of the Bondi model]{McNamaraNulsen2012}. Alternatively, these results could be evidence that jets can extract spin energy from the BH via the Blandford-Znajek mechanism \citep{BlandfordZnajek1977}, which allows efficiencies over $100\%$ \citep{TchekhovskoyEtAl2011, TchekhovskoyEtAl2012, McKinneyEtAl2012}. In any case, considering the lobe inflation process in more detail, as highlighted by \citet{McNamaraNulsen2007}, under the assumption that the jet kinetic energy is efficiently thermalised and the lobe inflation proceeds in pressure equilibrium with the ICM, the total energy needed to inflate the lobe is given by the lobe enthalpy,
\begin{equation}
    H = E_{\rm lobe} + PV = \frac{\gamma}{\gamma-1}PV,
\label{eq:enthalpy}
\end{equation}
where $\gamma$ is the adiabatic index of the lobe gas, $E_{\rm lobe}$ is the internal energy of the lobe and $PV$ is the product of the lobe pressure and volume. The resulting ratio of injected jet energy to lobe energy to work done is $E_{\rm jet}$:$E_{\rm lobe}$:$PV=1$:$1/\gamma$:$(\gamma - 1)/\gamma$, which for a non-relativistic gas ($\gamma = 5/3$, which is commonly assumed in simulations) suggests the lobes should retain $60\%$ of the injected energy with the remaining $40\%$ going into the ICM via work done. However, whether such behaviour is seen in reality depends on what dominates the lobes (e.g., relativistic or non-relativistic particles, magnetic fields etc) and hence determines the effective value of $\gamma$, and on the suitability of assuming that lobes are inflated in pressure equilibrium with their environment, i.e. how explosive the feedback is. With respect to the latter point, \citet{TangChurazov2017} performed idealised simulations of spherically symmetric AGN feedback events with varied energies and duration, finding that shorter, more explosive, injection events result in larger fractions of energy (up to $\sim 88\%$) going into shocks, while for longer, gentler, injection events the fraction of injected energy ending up in shocks is negligible. As such, the lobe enthalpy varies from $\sim 0$ for instantaneous injection up to roughly the injected energy for infinitely long injection episodes. Hydrodynamic simulations of jets in a cluster environment performed by \citet{HardcastleKrause2013} found that while the lobe energy is dominated by the thermal component, there are non-negligible kinetic and potential energy contributions. They additionally find that the ratio of lobe energy to energy gain by the ICM achieves a roughly constant value of $\sim$~one after an initial lobe inflation phase where the ratio can be larger, highlighting the non-negligible kinetic energy content of the lobe and the fact that they can be over-pressured compared to the ICM. Follow-up work of \citet{EnglishEtAl2016} presented relativistic jet simulations both with and without MHD for a range of jet velocities and powers. All jets studied inflate lobes that experience an initial phase of being over-pressured (up to a factor $\sim5$) before lobe pressures rapidly decline, and while low-power jets tend to come into rough pressure equilibrium with the ICM, lobe pressure ratios for their highest power jets never drop below $\sim 2$ and begin to increase again as the lobes move to larger radii. They additionally find that the shock-to-lobe energy ratios attain a roughly constant value of $\sim 1.5$, consistent with the results of earlier MHD simulations by \citet{HardcastleKrause2014} and slightly higher than values found in pure hydrodynamic simulations of \citet{HardcastleKrause2013}. Other jet simulations have additionally found ratios of $\gtrsim 1$, with varying levels of lobe kinetic energy, for both purely hydrodynamical simulations \citep{BourneSijacki2017, GuoEtAl2018, BourneEtAl2019, BourneSijacki2021, HuskoLacey2023Method} and those including magnetic fields \citep{HardcastleKrause2014, WeinbergerEtAl2017, EhlertEtAl2018}. \citet{BourneSijacki2021} find that due to the lobes initially being over-pressured, the instantaneous $PV$ and lobe enthalpy (calculated from equation \ref{eq:enthalpy}) underestimate the integrated $PdV$ work done during lobe inflation and total jet energy, respectively. \citet{PeruchoEtAl2017} additionally highlight the importance of modelling relativistic jets, which can result in larger lobe pressures by at least a factor of two. 

\subsubsection{The Lobe-ICM Interaction}
\label{sec:lobe_icm}

\begin{figure}
\includegraphics[width=13.5cm]{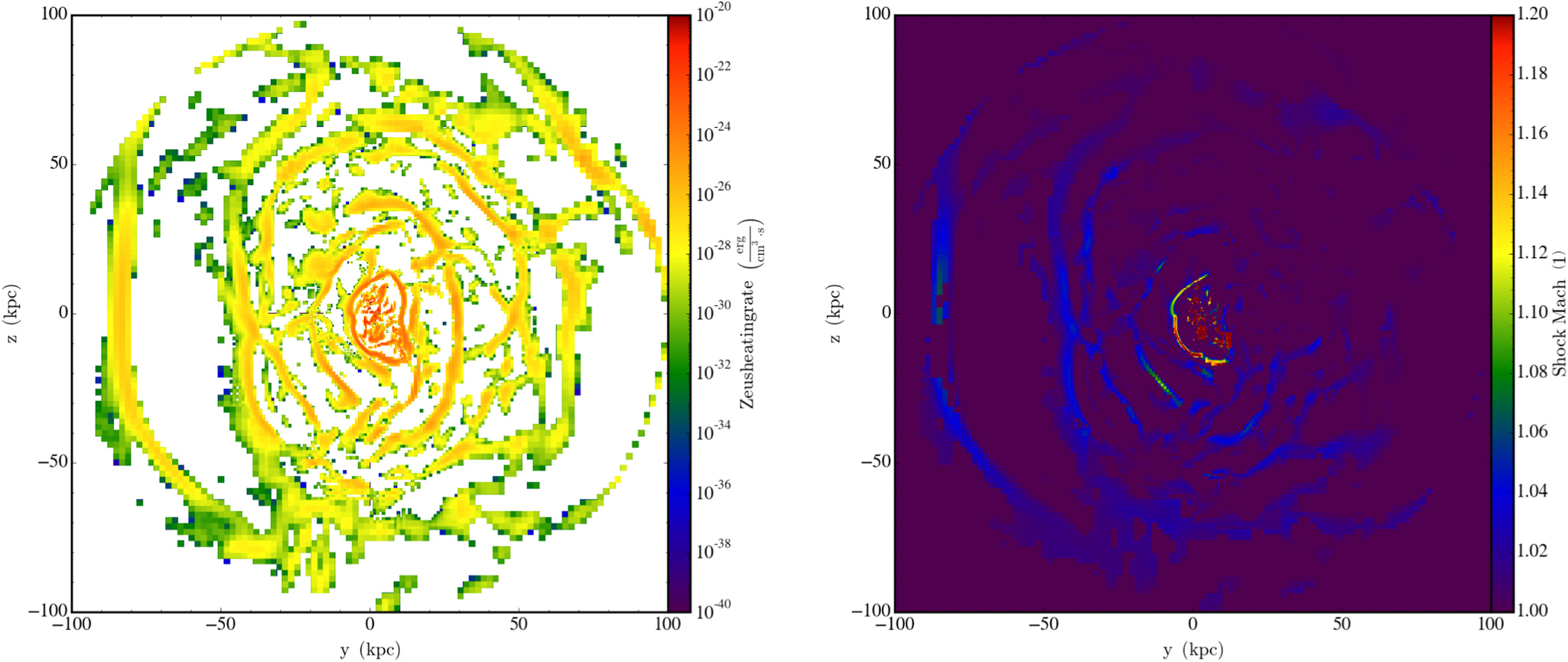}
\caption{An illustration of jet-driven shocks within the central $100$~kpc of a simulated cluster. The simulation includes self-regulated AGN feedback assuming jet power coupled to SMBH accretion of cold gas with an efficiency parameter of $\epsilon = 1\%$ (as defined in equation \ref{eq:mdot}). The jets are assumed to have a small angle precession ($\theta = 0.15$) with a 10 Myr period around the $z$-axis in the simulation. The left panel shows the energy dissipation rate while the right-hand panel shows shock Mach numbers, illustrating that stronger shocks and hence higher dissipation rates are seen closer to the jet but that overall the shocks are typically quite weak. (Figure 1 from \citet{LiEtAl2017}, \copyright AAS, reproduced with permission)}
\label{fig:li_shocks}
\end{figure} 

While it is clear that shocks play a critical role in lobe inflation, they can also make an important contribution to heating the ICM. While the jet itself drives a bow shock into the ICM, the initial rapid expansion of the jet lobes can also drive lateral shocks. As the lobe expansion slows the shocks become weaker and potentially broaden into sound waves, providing a mechanism through which energy can be communicated isotropically to the ICM \citep[e.g.][]{ReynoldsEtAl2001, BruggenEtAl2007, PrasadEtal2015, YangReynolds2016Hydro, BourneSijacki2017, LiEtAl2017, WeinbergerEtAl2017, GuoEtAl2018, MartizziEtAl2019, HuskoEtAl2022}. It is also worth noting that, similar to typical observations of jet feedback in galaxy clusters \citep{FabianEtAl2006, FormanEtAl2007, CrostonEtAl2011, SandersEtAl2016}, many modern simulations find that shocks are often not very strong \citep[e.g.][]{BruggenEtAl2007, SternbergSoker2009, LiBryan2014,  LiEtAl2017, BourneSijacki2021, EhlertEtAl2018, HuskoLacey2023Method}. \citet{LiEtAl2017} explicitly tracked the numerical dissipation within shock cells and found that heating due to shocks exceeded that due to turbulence by $\sim$~an order of magnitude and, as shown in Figure~\ref{fig:li_shocks} taken from \citet{LiEtAl2017}, that the vast majority of shocks were weak. Other works such as \citet{YangReynolds2016Hydro} and \citet{MartizziEtAl2019} estimate the dissipation rates from the expected entropy jump for weak shocks
\begin{equation}
    ds = \simeq \frac{2\gamma k_{\rm B}}{3(\gamma+1)^{2}\mu m_{\rm H}}(\mathcal{M}^{2}-1)^{3},
\end{equation}
finding that strong shocks can play an important role within the jet cone while weak shocks can dominate heating outside the cones, however, \citet{YangReynolds2016Hydro} find that weak shocks alone are unable to completely offset cooling. \citet{BourneSijacki2021}, who perform simulations in a cosmologically evolved cluster, make use of more conservative criteria to define shocks to avoid misclassification due to galaxy formation processes \citep[see][for details]{SchaalEtAl2015, SchaalEtAl2016}. They also find that weak shocks are an important component of the energy budget, albeit with somewhat lower levels of direct dissipation and instead highlight the importance of compressive heating at shock fronts.

As well as shocks, simulations also exhibit sound/compression waves driven by lobe expansion. These sound waves ``detach'' from the lobes once the expansion speed drops below the ICM sound speed and have the potential to travel isotropically to large distances \citep{RuszkowskiEtAl2004a, RuszkowskiEtAl2004b, SternbergSoker2009, DuboisEtAl2010, BourneSijacki2017, BourneSijacki2021, GuoEtAl2018, HuskoEtAl2022}. \citet{BambicReynolds2019} recently presented a detailed study of sound wave production by momentum-driven jets in an ICM atmosphere (see also Section~\ref{sec:soundwave} and Figure~\ref{fig:bambic_soundwaves}), finding that constructive interference can lead to $\gtrsim 25\%$ of the jet energy being converted to sound waves, exceeding the level expected from spherical blast waves ($\sim 10\%$) \citep{TangChurazov2017}. The rate of heating and volume over which sound waves are energetically important, though, depend on the microphysics of the ICM (see Section~\ref{sec:soundwave}) that dictates the cluster viscosity \citep{RuszkowskiEtAl2004a, RuszkowskiEtAl2004b, SijackiSpringel2006b, ZweibelEtAl2018}.

Several early simulations assumed that lobes can be simply modelled as low-density bubbles of hot gas that rise buoyantly through the cluster atmosphere \citep{ChurazovEtAl2001, BruggenKaiser2001, QuilisEtAl2001, BruggenEtAl2002, DallaVechiaEtAl2004}. However, in omitting the inflation phase these simulations miss the potential impact that the jet momentum has on bow shock production and the dynamics of the lobes, which may, at least initially, propagate on timescales shorter than the buoyant rise time \citep{OmmaEtAl2004, BourneSijacki2017, EhlertEtAl2018, EnglishEtAl2019}. Observations of clusters can exhibit multiple old cavities that go out to large radii \citep{FormanEtAl2007, DiehlEtAl2008, FabianEtAl2011, RandallEtAl2011, RandallEtAl2015}. Early simulations of bubbles were found to be disrupted by Rayleigh-Taylor and Kelvin-Helmholtz instabilities on relatively short timescales unless additional physics such as physical viscosity \citep{ReynoldsEtAl2005, SijackiSpringel2006b} or magnetic fields \citep{JonesEtAl2005, Dursi2007, RuszkowskiEtal2007, DursiPfrommer2008, OneillEtAl2009} were included. However, \citet{SternbergSoker2008} showed that the formation of a dense layer of gas around jet-inflated lobes can inhibit the development of such instabilities, thus allowing bubbles to travel further and live longer. This being said, simulations of wide jets (large opening angle) have found that vortices can develop that effectively mix lobe material with the ICM \citep{GilkisSoker2012, HillelSoker2016, HillelSoker2017}, with \citet{HillelSoker2016} finding that it is more effective than both turbulence and shock heating \citep[see also,][]{YangReynolds2016Hydro, LiEtAl2017}. On the other hand, simulations that make use of narrow jets ($0^{\circ}$ opening angle) find that mixing of jet material into the ICM via Kelvin-Helmholtz instabilities is sub-dominant \citep{BourneSijacki2017, WeinbergerEtAl2017, MartizziEtAl2019} and that the level of mixing reduces with increasing jet power \citep{EhlertEtAl2018, BourneSijacki2021}. \citet{BourneSijacki2017} highlight the impact of resolution on mixing, finding that lower resolution results in increased levels of mixing, citing the existence of larger Kelvin-Helmholtz vortices in this regime, and additionally highlight that appropriate refinement of the jet lobes is required to avoid excessive numerical mixing \citep[see also,][]{WeinbergerEtAl2017}. \citet{MartizziEtAl2019} highlight that the level of mixing can depend upon the hydro-solver implemented, which can impact how well mixing instabilities are resolved \citep[see also,][]{OgiyaEtAl2018}. Further, the inclusion of magnetic fields has the potential to inhibit mixing \citep[e.g.,][]{JonesEtAl2005, Dursi2007, RuszkowskiEtal2007, DursiPfrommer2008, OneillEtAl2009, BambicEtAl2018, WeinbergerEtAl2017, EhlertEtAl2018}. In addition, the ability to effectively mix lobe and ICM material depends on the constitution of the lobe material \citep{YangEtAl2019} and the plasma properties of the ICM (see Section~\ref{sec:viscosity}).

Observations of galaxy clusters can show multiple cavities at a range of radii representing different generations of jet lobes that have risen buoyantly through the ICM \citep{DunnEtAl2006, FormanEtAl2007, DiehlEtAl2008, FabianEtAl2011, RandallEtAl2011}. Such rising bubbles can displace significant amounts of ICM material and increase its gravitational potential energy \citep{HardcastleKrause2013, HardcastleKrause2014, WeinbergerEtAl2017, YangEtAl2019, BourneSijacki2021, HuskoLacey2023Interplay}. It was argued early on that displaced material would fall back behind a rising cavity, driving turbulence and ultimately converting its potential energy into heat, a process known as cavity heating \citep{ChurazovEtAl2001, ChurazovEtAl2002, BirzanEtAl2004}. \citet{YangReynolds2016Hydro} point out that the exact heating rate is dependent on the ICM viscosity and find that in any case kinetic energy within the wake, which would be the source of cavity heating, accounts for only a small fraction of the injected energy. Instead, \citet{YangReynolds2016Hydro} find that clusters can exhibit large circulation patterns whereby gas is lifted out of the cluster core and effectively heated by shocks and mixing within the jet cone, before flowing back to the cluster core. The role of jet-driven gas circulation in galaxy clusters has been highlighted by several works \citep[e.g.,][]{HillelSoker2016, SokerEtAl2016, GuoEtAl2018, ChenEtAl2019} with \citet{ChenEtAl2019} suggesting that if low entropy material can effectively mix with higher entropy material at large radii before returning to the cluster core, this process can act somewhat akin to a heat pump. Other simulations have found that the rising bubbles can lift low entropy material to larger radii, which can result in the formation of cold filaments and clumps, either through direct uplift or by stimulating condensation \citep[e.g.,][]{RevazEtAl2008, GaspariEtAl2012, LiBryan2014, BrighentiEtAl2015, GuoEtAl2018, BeckmannEtAl2019, WangEtAl2019, WangEtAl2020, HuskoLacey2023Interplay, HuskoEtAl2022, ZhangEtAl2022}. Such material can display properties similar to cold gas observed in galaxy clusters \citep[e.g.][]{SalomeEtAl2006, McDonaldEtAl2012, HamerEtAl2014, HamerEtAl2016, FabianEtAl2016, RussellEtAl2017}. Recent MHD simulations show that efficient coupling, facilitated by magnetic fields, can promote angular momentum transfer between hot gas stirred by jets and cold material, which along with the effects of magnetic braking can have a critical impact on cold gas angular momentum \citep{WangEtAl2020, WangEtAl2021, EhlertEtAl2022}. \citet{EhlertEtAl2022} show that this in turn has an impact on cold gas morphology as it promotes the formation of transient cold discs and radial filaments as opposed to the long-lived discs seen when magnetic fields are not included. This cold gas can play an important role in the jet life-cycle as a source of future fuel, with many simulations that implement cold accretion models to power momentum-driven jets \citep[e.g.,][]{GaspariEtAl2012, GaspariEtAl2013, LiBryan2014, LiEtAl2017, YangReynolds2016Hydro, PrasadEtal2015, MeeceEtAl2017} finding good agreement between simulated and observed CC cluster properties.

Assuming the lobe lifetimes are sufficiently long, they can excite g-modes within the cluster atmosphere \citep{ChurazovEtAl2002, OmmaEtAl2004, ZhangEtAl2018}. It has been argued that such modes can decay into volume-filling turbulence, which then dissipates to heat the ICM \citep{ZhuravlevaEtAl2014, ZhuravlevaEtAl2016}. However, idealised hydrodynamic simulations performed by \citet{ReynoldsEtAl2015} found that only a small fraction ($\lesssim 10\%$) of the injected energy ends up as turbulence \citep[similar results were also found when including magnetic draping,][]{BambicEtAl2018}. Other simulations that specifically include jets have found that the level of turbulence produced is subdominant compared to the total injected jet energy and often located in the jet vicinity itself \citep[e.g,][]{YangReynolds2016Hydro, BourneSijacki2017,  HillelSoker2017Hitomi, WeinbergerEtAl2017, LiEtAl2017, PrasadEtAl2018}. On the other hand, \citet{MartizziEtAl2019} found that jet-driven turbulence can dominate the heating budget in the central regions of the cluster, although this is only when the jet is active, and its contribution becomes negligible at large radii. The inclusion of additional physical processes may also affect the turbulent contribution to the heating budget. Simulations of \citet{BeckmannEtAl2019}, that include spin-driven jets find that continuous evolution of the jet direction can drive turbulence over a larger volume, while recent results from \citet{WangEtAl2021} found that the inclusion of magnetic fields can promote jet-driven ICM turbulence due to tighter coupling between jets and the ICM. Some simulations that include ``cluster weather'' \citep{BourneSijacki2017, LauEtAl2017, EhlertEtAl2021}, find that large-scale turbulence and bulk motions, such as that driven by orbiting substructures or cosmic accretion, are required in combination with AGN feedback to match both the line of sight velocities and velocity dispersions measured from Hitomi observations of the Perseus cluster \citep{Hitomi2016, HitomiEtAl2018}. Specifically, while simulated AGN feedback could produce the required levels of velocity dispersion, it was unable to produce other features such as the large velocity shear. Although turbulence may be energetically unimportant, it can play other roles. While the motions of rising lobes can move metals and create elongated distributions \citep[e.g.,][]{RoedigerEtAl2007, GaspariEtAl2011a, GaspariEtAl2011b, DuanEtAl2018}, jet induced turbulence has been found to additionally effectively redistribute the metals  \citep{GaspariEtAl2011a, GaspariEtAl2011b}. Additionally, \citet{GaspariEtAl2013} showed that subsonic turbulence can promote thermal instability and provide a supply of low angular momentum cold gas to feed the central black hole through ``chaotic cold accretion'' \citep[see also,][]{McCourtEtAl2012, GaspariEtAl2015, GaspariEtAl2017, WangEtAl2019, WangEtAl2020}.

\subsection{The Role of Environment and Cluster Weather}
\label{sec:macro_environment}

Having focused on the general macroscopic evolution of jet lobes in a galaxy cluster, in this section we consider the role that the environment can play in shaping lobe properties and how they interact with the ICM. At the simplest level, the ICM can be modelled as a smooth, hydrostatic atmosphere with some radially dependent density and temperature profiles, with both the absolute density \citep[e.g.,][]{YatesEtAl2018, PeruchoEtAl2022} and profile slope \citep[e.g.,][]{HardcastleKrause2013, HardcastleKrause2014, EnglishEtAl2019, MassagliaEtAl2022, MandalEtAl2022} being able to impact jet morphology, propagation and radio emission properties. \citet{YatesEtAl2021} further showed that offsetting the jet launch location from the cluster centre and varying the jet direction can impact the jet properties, with the jets encountering a higher density medium typically being shorter and having hot spots with higher surface brightness. Building on this, the ICM is not a simple single-phase gas but rather contains both a hot X-ray emitting plasma as well as cold filaments and disc structures \citep[e.g.,][]{SalomeEtAl2006, McDonaldEtAl2012, HamerEtAl2014, HamerEtAl2016, FabianEtAl2016, RussellEtAl2017}. Small-scale simulations ($\lesssim 1$~kpc) of jets show that their interaction with the cold dense clouds or discs can impact jet morphology and dynamics as well as how effectively the jet couples to the surrounding medium \citep[e.g.,][]{WagnerBicknell2011, WagnerEtAl2012, MukherjeeEtAl2018, MukherjeeEtAl2021, TalbotEtAl2021, TalbotEtAl2022}. While in cluster scale simulations, the interaction of jets with cold gas can help thermalise the jets \citep{PrasadEtal2015, MeeceEtAl2017}, inhibit their propagation \citep{BourneEtAl2019} and in the case of light jets deflect them, making their interaction with the ICM more isotropic \citep{EhlertEtAl2022}.

Observations suggest that the ICM is turbulent \citep[e.g.,][]{SchueckerEtAl2004, SandersEtAl2011, ZhuravlevaEtAl2012, PintoEtAl2015, WalkerEtAl2015, Hitomi2016, HitomiEtAl2018, HofmannEtAl2016} and while some simulations suggest that jets can drive local turbulence \citep{YangReynolds2016Hydro, MartizziEtAl2019, GaspariEtAl2012, GaspariEtAl2013, PrasadEtAl2018, YangEtAl2019, WangEtAl2021, EhlertEtAl2022}, other processes such as those arising from structure formation and sloshing are expected to drive widespread turbulence and bulk motions within the ICM \citep[e.g.,][]{RuszkowskiEtAl2011, VazzaEtAl2012, VazzaEtAl2017, ZuHoneEtAl2013, BourneSijacki2017, BourneEtAl2019, BourneSijacki2021, LauEtAl2017, Valdarnini2019, BennettSijacki2022}, while tangled magnetic fields may also drive MHD turbulence \citep{EhlertEtAl2018, EhlertEtAl2019, EhlertEtAl2021}. Such ``cluster weather'' could play an important role in lobe dynamics, with a number of studies attempting to mimic its effects by applying small perturbations to spherically symmetric environments. For example \citet{Krause2005} included small density perturbations to a King atmosphere, finding that this results in differences in back-flow locations and small asymmetries in jet lengths. \citet{OneillJones2010} performed MHD simulations of jets in a cluster environment containing a tangled magnetic field and an ambient medium on which fluctuations ($\pm 10\%$) in density have been superimposed to match observed ICM pressure variations. However, while these fluctuations could perturb the jets, they were insufficient to significantly alter the jet dynamics and energy only primarily couples to the ICM within the jet cone. On the other hand, some works instead induce ICM gas motions, for example \citet{BourneSijacki2017} used hydrodynamic simulations to consider the long-term evolution of jets and their lobes in a cluster stirred by orbiting substructures, finding that the resulting turbulent and bulk motions can deflect the lobes and increase mixing of lobe and ICM material when compared to a hydrostatic atmosphere. \citet{EhlertEtAl2018} study the dynamics of jets in an ICM atmosphere including a turbulent magnetic field, finding that while magnetic draping can suppress instabilities the presence of turbulence can reduce the effects of this draping and in the buoyant rise phase disrupt the jet lobes, while \citet{EhlertEtAl2022} find that the turbulence can deflect bubbles, assuming the jet is sufficiently light and provide a more isotropic distribution of lobe positions, akin to observations of multiple generations of lobes in some galaxy clusters \citep[e.g.,][]{FormanEtAl2007, DiehlEtAl2008, ChonEtAl2012, FabianEtAl2011, UbertosiEtAl2021}. However, as we discuss later in this section, only a handful of studies include high resolution jets in cosmologically evolved cluster environments \citep{HeinzEtAl2006, MorsonyEtAl2010, MendygralEtAl2012, BourneEtAl2019, BourneSijacki2021, YatesEtAl2023}, and, to the best of our knowledge, just one includes self consistently evolved magnetic fields \citep{MendygralEtAl2012}.

Observed radio jets can exhibit bent morphologies \citep[e.g.][]{PatnaikEtAl1986, DehghanEtAl2014, OBrienEtAl2018}. Ram pressure due to crosswinds, arising from relative motion between jets and the ICM, is expected to lead to such bending and result in the formation of narrow-angle-tailed (NAT) and wide-angle-tailed (WAT) radio galaxies \citep{MorsonyEtAl2013, GanEtAl2017, JonesEtAl2017, OneillEtAl2019a, OdeaBaum2023}. Further to this, cluster-merger-driven shocks can also impact jet morphology. \citet{GanEtAl2017} compared hydrodynamic and magnetically dominated jets when subjected to crosswinds and shocks to mimic cluster weather, finding that while the hydrodynamic jets were unable to survive, the magnetic jets can be significantly bent in the presence of a cross-wind but not solely by a shock front. \citet{OneillEtAl2019b} simulated a shock front passing a NAT radio galaxy, finding that the shock action compresses the jets and that turbulence increases in the jet tails, which in turn amplifies the magnetic fields and promotes mixing. \citet{NoltingsEtAl2019a, NoltingEtAl2019b} extended this work to consider the interaction of merger-like shock fronts impacting generic jet pairs. Shocks moving parallel to the jet axis can significantly affect the jet dynamics, with the post-shock flow potentially impeding or reversing the jet and destroying the cocoon \citep{NoltingsEtAl2019a}. Similarly, when the jet axis is in the plane of the shock front, the jet lobes are disrupted and in the case of active jets, the action of the shock bends them to form a NAT radio galaxy \citep{NoltingEtAl2019b}. 

To self-consistently capture large-scale turbulence driven by structure formation, full cosmological simulations are necessary \citep[e.g.,][]{GenelEtAl2014, SchayeEtAl2015, DuboisEtAl2016, McCarthyEtAl2017, HendenEtAl2018, SpringelEtAl2018}. Models have been developed to capture the low accretion state associated with jet feedback, for example, by adding bubbles of hot gas to mimic radio lobes \citep[e.g.,][]{SijackiSpringel2006a, SijackiEtAl2007, SijackiEtAl2015}, or driving bi-polar kinetic outflows \citep[e.g.,][]{DuboisEtAl2010, DuboisEtAl2012, DaveEtAl2019}, that can regulate cooling and star formation within galaxy clusters. However, as discussed in Section~\ref{sec:CosmoSims}, to capture representative volumes of the Universe over cosmic timescales, these models by necessity have to make simplifying assumptions (e.g. low velocities, high mass loading and/or hydrodynamic decoupling) and sacrifice resolution in the injection region and/or jet lobes. While the focus of such simulations is on capturing the global effects of the feedback on resolvable scales as opposed to studying lobe inflation and subsequent interaction with the ICM in intricate detail, they are still able to reproduce observational features, such as X-ray rims and cavities \citep{SijackiSpringel2006a, SijackiEtAl2007, DuboisEtAl2010}, and the effects of cluster weather displacing lobes and promoting mixing \citep{SijackiEtAl2008} that are seen in the high resolutions simulations we discuss below. Additionally, cosmological simulations of galaxy clusters have been used, for example by \citet{VazzaEtAl2021}, to study how cluster dynamics can impact the evolution of relativistic electron populations and magnetic fields produced by radio galaxies \citep[see also,][]{VazzaEtAl2023a, VazzaEtAl2023b}, finding that cluster motions interact with radio lobes, affecting their dynamics, constituent magnetic fields, and radio properties.

\citet{BruggenEtAl2005} presented some of the earliest simulations to consider radio mode feedback in a cosmologically evolved cluster by injecting thermal energy off-centre from the cluster core to mimic the inflation of jet lobes in a dynamic environment. However, \citet{HeinzEtAl2006} presented the first example of high-resolution bipolar jets in a dynamic cluster environment, using non-radiative, hydrodynamic simulations that included $10^{46}$~erg~s$^{-1}$ jets in the centre of a cosmologically evolved, $7\times10^{14}$~M$_{\odot}$ cluster\footnote{While the simulations performed by \citet{HeinzEtAl2006} were non-radiative, the simulations used to produce the initial conditions did include radiative cooling and star formation \citep{SpringelEtAl2001}.}. ICM motions were able to effectively move the lobes and redistribute material, helping to make the energy injection into the ICM more isotropic and replenish higher-density material at small radii for later jets to couple to. Mock radio and X-ray images exhibited lobe and jet morphologies and features as seen in real observations. A further parameter study using the same cluster but varying jet power and lifetime was performed by \citet{MorsonyEtAl2010}. As in \citet{HeinzEtAl2006}, the ICM motions effectively displace, disrupt and mix the jet lobes, with more powerful jets being able to affect the ICM at larger radii. Interestingly, \citet{MorsonyEtAl2010} found that single AGN outbursts can potentially be realised as multiple independent lobe structures that could otherwise be interpreted as multiple jet outbursts. MHD simulations performed by \citet{MendygralEtAl2012} studied $6\times10^{44}$~erg~s$^{-1}$ jets in a non-radiative cosmologically evolved cluster that hadn't undergone any recent mergers. They found that the magnetic field lines are swept up by and drape over the inflated lobes and that despite the dynamically relaxed nature of the system, ICM motions can still impact the lobe morphologies and dynamics, with mock radio images exhibiting morphologies akin to WAT radio galaxies with bent jets.

\begin{figure}
\includegraphics[width=13.5cm]{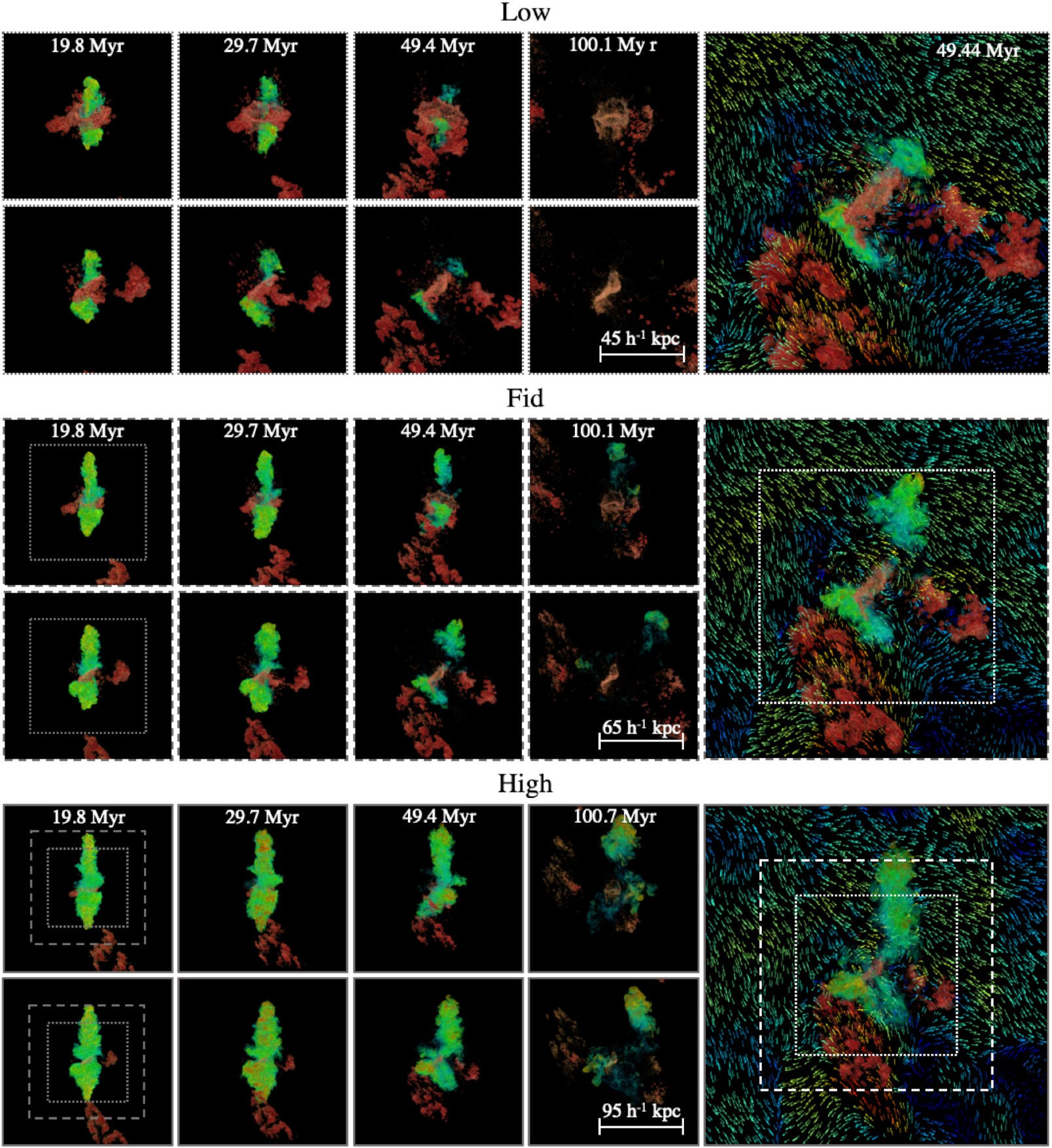}
\caption{Volume rendering of jet lobes (green) and cold material (red) within a cosmologically evolved galaxy cluster. The top, middle and bottom rows show low, medium (labelled ``Fid'') and high-power jets, respectively. The small panels show the evolution of the jet lobes for two different viewing angles (rotated by $90^{\circ}$ about the $z-$axis with respect to each other), while the large panels additionally overlay the gas velocity field. Overall, jet lobes can be displaced, disrupted and mixed by cluster weather and cold structures, with lower-power jets being more susceptible. (Figure 2 from \citet{BourneSijacki2021}, CC BY 4.0)}
\label{fig:bourne_weather}
\end{figure} 

Later work by \citet{BourneEtAl2019} presented radiative hydrodynamic moving-mesh simulations of a $4\times10^{44}$~erg~s$^{-1}$ jet pair in a low-redshift cosmologically evolved cluster that exhibited a massive cold central disc structure. Mock X-ray images show classic cavities and bright rim structures, with both the cold gas and cluster weather affecting the lobe structure and location, with the latter acting to displace lobes from their initial trajectory. Shocks are initially driven into the ICM during lobe inflation and similar to results in hydrostatic atmospheres \citep[e.g.,][]{HardcastleKrause2013, EnglishEtAl2016, BourneSijacki2017, WeinbergerEtAl2017}, $\sim 40\%$ of the jet energy goes into $PdV$ work. However, at later times the cluster weather can effectively mix jet and ICM material, with the aid of an in-falling substructure that directly impacts one of the lobes. \citet{BourneSijacki2021} extended this work to study different jet powers and perform a more detailed analysis of ICM heating. Figure~\ref{fig:bourne_weather}, taken from their paper illustrates the evolution of different power jets and the lobes that they inflate when subjected to cluster weather. Overall, they found that weak shocks and weather-aided mixing dominate the heating budget, and that while high-power jets drove stronger shocks that were more efficient at thermalising energy, lower-power jets and the lobes they inflate simply move material around and were more susceptible to weather-aided mixing \citep[see also,][]{EhlertEtAl2018}. As well as recovering cavities and rims in mock X-ray observations, mock radio images show morphologies ranging from FR-I at low power to FR-II at high power \citep[see also,][]{MassagliaEtAl2016, EhlertEtAl2018}. 

Both FR-I and FR-II jet morphologies were also realised in the simulations of \citet{YatesEtAl2023}, who presented the initial results of the CosmoDRAGoN project, the first simulations to study relativistic jets in cosmologically evolved clusters. The project performed a parameter study by simulating fixed power jets in a cosmologically evolved cluster (taken from the 300 Project \citep{CuiEtAl2018}) varying both velocity (from non-relativistic to relativistic) and jet opening angle. They produce mock radio surface brightness maps which illustrate both core-brightened FR-I and edge-brightened FR-II sources. As in previous works performed in idealised cluster environments \citep{KrauseEtAl2012}, they recover the result that FR-I sources can arise from wider jets. They further find that slower jets transition from being FR-II to FR-I sources sooner and additionally highlight that even their most powerful jets eventually transitions to FR-I provided it has a wide enough opening angle. Although to a lesser extent than found in some of the simulations discussed above, \citet{YatesEtAl2023} also find that cluster weather can affect lobe morphology, particularly low power (FR-I) sources once the jet has been switched off and the lobe enters the buoyant rise phase \citep[see also,][]{BourneSijacki2021}. 

\citet{VernaleoReynolds2006} raised the problem that if jet direction remains unchanged, the low-density path cleared by an early jet episode remains, and later feedback events can escape largely uninhibited, resulting in inefficient coupling between the jet and the ICM. As discussed in Section~\ref{sec:morphology_direction_energetics}, various approaches have been taken to improve coupling between the jets and ICM, and to distribute energy more isotropically (e.g. jet precession/reorientation, large opening angles, light jets). Processes driving both precession and jet re-orientation are still somewhat uncertain and whether or not accretion discs can do this on short timescales is debated \citep{NixonEtAl2013}, with SMBH binaries being an alternative method to drive precession \citep{BegelmanEtAl1980, KrauseEtAl2019}. In any case, the simulations discussed in this section, which include dynamic cluster environments, provide an alternative natural relief to the problem, by removing the low-density channels, redistributing the lobes, and promoting mixing and distributing the energy isotropically. In displacing lobes, cluster weather may also provide a mechanism to explain observations of multiple lobe generations having quite different trajectories \citep{DunnEtAl2006, FormanEtAl2007, DiehlEtAl2008, FabianEtAl2011, FabianEtAl2022}.

As already mentioned in Section~\ref{sec:bubbles_shocks_waves}, magnetic fields can play a potentially important role in lobe dynamics and interactions. Observations of galaxy clusters show they have an extended magnetic field with a strength of a few to tens~$\mu$G and plasma $\beta\sim100$ \citep{CarilliEtAl2002, FerettiEtAl2012}. Simulations have shown that bubbles moving above the Alfv\'en speed can ``sweep up'' field lines and form a draping layer, in which the magnetic field is amplified. This can suppress instabilities along the direction of the field lines, reduce mixing and slow the bubble due to magnetic tension \citep[e.g.,][]{JonesEtAl2005, Dursi2007, RuszkowskiEtal2007, DursiPfrommer2008, OneillEtAl2009, BambicEtAl2018, EhlertEtAl2018}. Provided the Alfv\'en speed in the draping layer is sufficiently large compared to the buoyancy or shear velocities, Rayleigh-Taylor and Kelvin-Helmholtz instabilities with wavelengths up to an order of magnitude larger than the layer thickness can be suppressed \citep{Dursi2007}. While the formation of the draping layer is found to require that the coherence scale of the magnetic field is larger than the bubble \citep{RuszkowskiEtal2007, OneillEtAl2009, PfrommerDursi2010}, \citet{OneillEtAl2009} also highlighted that the dynamical effect of the draping layer can be reduced by tangled magnetic fields and that effective amplification requires that the scale on which the fields are tangled is greater than the bubble. Further, the existence of turbulence in the ambient medium can perturb the lobe surface, inhibit magnetic draping and potentially reduce its effectiveness at suppressing mixing \citep{EhlertEtAl2018}. 3D simulations of jets in tangled magnetic fields have also shown that field lines carried with the jet lobes can align as they are stretched and amplified within the wake of rising bubbles \citep{RuszkowskiEtal2007, OneillJones2010, MendygralEtAl2012, EhlertEtAl2018}. As mentioned in Section~\ref{sec:lobe_icm} already, ICM magnetic fields can promote coupling between the jet and ambient medium leading to the generation of ICM turbulence \citep{WangEtAl2021}, while cold gas angular momentum can also be affected due to magnetic braking and increased coupling between the hot and cold gas phases \citep{WangEtAl2020, WangEtAl2021, EhlertEtAl2022}, which directly impacts cold gas morphology and has implications for SMBH feeding. On the other hand, \citet{BambicEtAl2018} found that the magnetic tension can inhibit g-modes decaying into volume-filling turbulence. Finally, it is becoming ever more important to include realistic magnetic fields in cluster simulations to explore microphysical processes such as anisotropic thermal conduction (Section~\ref{sec:conduction}), anisotropic viscosity (Section~\ref{sec:viscosity}) and CR transport mechanisms (Section~\ref{sec:cr}).

\subsection{Open Questions and Future Opportunities}
\label{sec:macro_future}

As is clear from the examples given in previous subsections, there is a burgeoning collection of jet models implemented in a range of codes currently used in the literature. While many of these models, which include self-regulated feedback, are now able to successfully prevent over-cooling within galaxy clusters, the mechanisms through which this is achieved are not necessarily agreed upon. Some processes such as the generation of weak shocks are common between simulations, albeit with different relative importance to the heating budget. On the other hand, processes such as mixing and the generation of turbulence appear to be dependent on the jet injection model, resolution and/or refinement techniques and physics included (e.g. MHD, viscosity, cluster weather, etc) and as such are not firmly settled. This being said, several works are now finding that the process is expected to be a gentle continuous one \citep{YangReynolds2016Hydro, MeeceEtAl2017, BourneSijacki2021, EhlertEtAl2022}. \citet{MartizziEtAl2019} showed that the Riemann solver used can affect the jet and lobe evolution depending on how diffusive the solver is and on how well small-scale physical processes, such as mixing, are captured. Additionally, the recent study of \citet{WeinbergerEtAl2022} argue that differences between codes, and the differences between model parameters within a code make bigger differences to results than having high resolution or numerical convergence. As such, while we can accept that jet feedback regulates the thermodynamics of the ICM, the exact mechanisms are not agreed upon and further constraints need to be placed on jet feedback models based on observations. This could be achieved by making more use of mock X-ray and radio images and comparing them to observations of lobe morphologies, distributions and lifetimes \citep[e.g.,][]{ShinEtAl2016}. Future X-ray missions with a larger effective area and hence higher signal-to-noise ratio, such as AXIS\footnote{\label{note:axis}\href{https://axis.astro.umd.edu/}{https://axis.astro.umd.edu/}}\citep{AXISWP2019}, could potentially reveal faint cavities at larger radii and increase the sample size of cavities for a more comprehensive statistical study.

The role of environment is still an open question, only a few works include high-resolution jet feedback in clusters with self-consistently driven weather \citep{HeinzEtAl2006, MorsonyEtAl2010, MendygralEtAl2012, BourneEtAl2019, BourneSijacki2021}, even fewer include cluster weather and magnetic fields \citep{MendygralEtAl2012}. A parameter space study of different cluster environments, with a range of dynamical states and over cosmologically interesting timescales needs to be considered to fully understand the role that cluster weather can play in displacing lobes, mixing lobe material and the role that magnetic fields can play in these process, i.e. to what levels of ICM turbulence are the fields still able to inhibit mixing? In relation to the dynamical state of the system, recent works suggested cluster mergers play an important role in driving the CC/NCC distribution \citep{BurnsEtAl2008, RasiaEtAl2015, HahnEtAl2017, ChadayammuriEtAl2021}, although other recent work found little difference in the dynamical state of CC versus NCC clusters \citep{BarnesEtAl2018}. As highlighted by \citet{BourneSijacki2021}, if CC clusters are more dynamically quiet, one could expect that lobes are more likely to survive, conversely if NCCs are dynamically active, it could be that lobes are more likely to be disrupted, such behaviour could additionally explain why lobes and bubbles are prevalent in CC clusters \citep{DunnFabian2006, Fabian2012}, although there is also an observational bias towards being able to detect cavities in CC vs NCC clusters due to higher X-ray photon counts in the former \citep[e.g.][]{PangouliaEtAl2014, HlavacekLarrondoEtAl2022}. The upcoming XRISM\footnote{\label{note:xrism}\href{https://xrism.isas.jaxa.jp/en/}{https://xrism.isas.jaxa.jp/en/}} mission \citep{XRISMWP2020}, a replacement for the ill-fated Hitomi\footnote{\href{https://www.isas.jaxa.jp/en/missions/spacecraft/past/hitomi}{https://www.isas.jaxa.jp/en/missions/spacecraft/past/hitomi}} mission, as well as more distant missions such as Athena\footnote{\label{note:athena}\href{https://www.the-athena-x-ray-observatory.eu/en}{https://www.the-athena-x-ray-observatory.eu/en}}\citep{AthenaWP2013}, will provide unprecedented measurements of the ICM velocity structure and turbulence. Such measurements, combined with observations of jet lobes should allow us to begin to address the interplay between cluster weather and AGN feedback. It is also possible that such observations, in combination with future simulations that include both ICM motions and models that track the evolution of SMBH/accretion disc angular momentum and jet direction, could play a role in untangling the apparent degeneracy between intrinsic jet direction and cluster weather in determining lobe spatial distributions.

The focus of this review is on the role of AGN feedback in the form of jets, often dubbed the ``radio mode'' \citep{Fabian2012}, which is expected to be the dominant mechanism for regulating heating and cooling in the ICM. This being said, accreting SMBHs can also release energy in the form of radiation and wide-angled winds that have the potential to impact their environment. Several models employed in cosmological simulations have been developed that invoke separate modes of feedback --- ``quasar'' vs ``radio'' --- during periods of high and low accretion (in terms of the Eddington rate), respectively, to differentiate between radiatively efficient vs radiatively inefficient accretion \citep[e.g.,][]{SijackiEtAl2007, DuboisEtAl2012, SijackiEtAl2015, WeinbergerEtAl2018, DaveEtAl2019}. However, as highlighted by \citet{QiuEtAl2019}, studies on the effects of radiative feedback in CCs are lacking. To begin to remedy this \citet{QiuEtAl2019} performed simulations invoking both kinetic outflows and radiative feedback in an idealised Perseus-like cluster. Further, while they invoke a transition accretion rate between radiatively efficient and inefficient modes, they allow kinetic feedback to occur simultaneously with radiative feedback at high accretion rates finding that to prevent over-cooling of the ICM kinetic outflows must be active across all accretion rates. Such works suggest there is ample opportunity to further study the role and impact of different feedback channels in shaping CC clusters. In addition, it has been argued by \citet{HardcastleCroston2020}, that the ``traditional'' paradigm of clearly separated quasar and radio modes distinguished by whether accretion is radiatively efficient or inefficient, and their role in galaxy formation is not so clear cut in reality, citing, for example, populations of high power radio jets occurring via radiatively efficient accretion at high Eddington ratios \citep[e.g.,][]{HineLongair1979, LaingEtAl1994, MingoEtAl2014, HardcastleCroston2020}. Additionally, simulations of radio-mode feedback typically do not explicitly implement different models for low-power/FR-I jets and high-power/FR-II jets. However, their environmental dependence \citep{ComerfordEtAl2020} and potential evolutionary sequence \citep{MacconiEtAl2020} could be used to constrain future models of accretion and/or feedback. As such, not only is there scope to consider distinct feedback regimes and how their efficiency couples to the accretion mode, but as in \citet{QiuEtAl2019} to further develop and explore models for determining how feedback should be channelled, potentially simultaneously, through different mechanisms (i.e. via radiation, winds and/or jets). Additionally, self-consistent lobe inflation, its influence on SMBH accretion, as well as the interaction of jets with the interstellar medium of the host galaxies require high-resolution simulations on galaxy scales \citep[e.g.,][]{AntonuccioDeloguSilk2010, WagnerBicknell2011, WagnerEtAl2012, MukherjeeEtAl2018, MukherjeeEtAl2021, TalbotEtAl2021, TalbotEtAl2022}. These simulations could provide valuable insights and inform future SMBH accretion and jet feedback models on cluster and cosmological scales.

Finally, as highlighted in the introduction, clusters sit at the crossroads of cosmology and astrophysics. In representing the high mass end of the halo mass function and the culmination of hierarchical structure formation, galaxy clusters have the potential to be used as probes of cosmology \citep{AllenEtAl2011, KravtsovBorgani2012}, with a number of current/future/proposed survey missions (e.g. DES\footnote{\href{https://www.darkenergysurvey.org/}{https://www.darkenergysurvey.org/}}\citep{DESWP2005}, SPT-3G\footnote{\href{https://www.anl.gov/hep/spt3g}{https://www.anl.gov/hep/spt3g}}\citep{SPT3GWP2014}, eRosita\footnote{\href{https://www.mpe.mpg.de/eROSITA}{https://www.mpe.mpg.de/eROSITA}}\citep{ERositaWP2012}, Advanced ACTPol\footnote{\href{https://act.princeton.edu/}{https://act.princeton.edu/}}\citep{ActPOLWP2016}, Euclid\footnote{\href{https://sci.esa.int/web/euclid}{https://sci.esa.int/web/euclid}}\citep{EuclidWP2011}, Vera C. Rubin Observatory's LSST\footnote{\href{https://rubinobservatory.org/}{https://rubinobservatory.org/}}\citep{LSSTWP2012}, Athena$^{\ref{note:athena}}$, SKAO\footnote{\href{https://www.skao.int/en}{https://www.skao.int/en}}\citep{SKAO2015}, Simons Observatory\footnote{\href{https://simonsobservatory.org/}{https://simonsobservatory.org/}}\citep{SimonsObsWP2019}, AXIS$^{\ref{note:axis}}$, CMB-S4\footnote{\href{https://cmb-s4.org/}{https://cmb-s4.org/}}\citep{CMBS42016}, CMB-HD\footnote{\href{https://cmb-hd.org/}{https://cmb-hd.org/}}\citep{CMBHDWP2019}, ngVLA\footnote{\href{https://ngvla.nrao.edu/}{https://ngvla.nrao.edu/}}\citep{NGVLAWP2019}) aiming to leverage this opportunity. In order to do this one typically needs to be able to accurately measure galaxy cluster masses, however, as their mass is dominated by dark matter, this cannot be done directly. One solution is to use cluster observables (e.g., X-rays properties, Sunyeav-Zel’dovich (SZ; \citep{SZ1972}) flux, radio emission, weak lensing) to estimate cluster masses, which requires a thorough understanding of the appropriate cluster mass scaling relations. Given that AGN feedback plays such a pivotal in shaping the ICM properties that give rise to these observables, it is imperative to continue developing cosmological simulations of structure formation that include robust modelling of jet feedback as well as additional baryonic physics, such as magnetic fields, CRs, conduction and viscosity, in order to fully understand cluster observable scaling relations, how they evolve with redshift, and understand any biases in derived masses \citep[see e.g.,][for recent examples]{BarnesEtAl2017a, HendenEtAl2019}. Furthermore, in redistributing material within galaxy groups and clusters, AGN feedback can directly impact the matter power spectrum, knowledge of which is necessary to derive cosmological parameters from weak lensing observations \citep[see discussions in][and associated references]{ChisariEtal2019, VanDaalenEtAl2020, AmonEfstathiou2022}. While existing cosmological simulations typically agree that feedback causes suppression of the matter power spectrum, the relevant levels and scales at which this occurs are model dependent \citep[e.g.,][]{VanDaalenEtAl2011, ChisariEtAl2018, ChisariEtal2019, SpringelEtAl2018, VanDaalenEtAl2020}, with ample scope remaining to constrain the relative importance of jet feedback in groups and clusters.

\section{Modelling the Microphysics}
\label{sec:micro}
Much of our understanding of AGN feedback in galaxy clusters has come from the extensive literature based on ideal hydrodynamic/MHD simulations reviewed in the previous section. However, some of the important physical processes, such as thermal conduction, viscosity, and CRs, have been largely neglected, and their effects on cluster feedback are relatively poorly understood. As mentioned in \S~\ref{sec:intro_importance}, these physical processes, including conduction and viscosity coefficients as well as the propagation of CRs, are determined by plasma physics that happen on microscopic scales compared to the sizes of galaxies and clusters, i.e., the ``microphysics''. In this section, we will review the impact of these microphysical mechanisms on AGN feedback in clusters. We will discuss the roles of CRs in Section~\ref{sec:cr} and the roles of ICM plasma physics in Section~\ref{sec:plasma}. 

\subsection{Roles of Cosmic Rays}
\label{sec:cr}

\subsubsection{Motivations for Considering Cosmic Rays}
\label{sec:cr_motivations}

% Growing recent literature about importance of CRs in galaxy and cluster formation in general, roles of CRs in clusters in general & gamma-ray limits

Effects of CRs on the formation and evolution of galaxies and galaxy clusters in the cosmological context have received growing recognition in recent years. In particular, state-of-the-art cosmological simulations have shown that baryonic feedback processes from stars and SMBHs play a vital role in reproducing the observed luminosity functions of galaxies \citep[e.g.,][]{DiMatteoEtAl2005, DuboisTeyssier2010, StinsonEtAl2013, LeBrunEtAl2014, SchayeEtAl2015, SijackiEtAl2015, DuboisEtAl2016, WeinbergerEtAl2018, MartinAlvarez2023}. For lower-mass galaxies, starburst-driven galactic outflows are key sources of feedback for suppressing SFRs; however, the physical mechanisms for launching galactic outflows had been elusive. Only recently have simulations shown that outflows with mass loading factors comparable to observed levels can be driven when the effects of CRs are included \citep{BoothEtAl2013, SalemBryan2014, RuszkowskiEtAl2017a}. These studies further showed that the properties of the galactic outflows, SFRs within the galaxies, as well as the circumgalactic medium (CGM), all sensitively depend on the detailed CR transport processes on microscopic scales \citep{ButskyEtAl2018,  JiEtAl2020, BuckEtAl2020}. These results point to the importance of CRs in the processes of structure formation, and the necessity for understanding how CRs propagate and interact with the magnetised medium. 

While on the scales of galaxies, CRs are in rough energy equipartition with thermal gas, magnetic fields, and turbulence, on the scales of galaxy clusters CRs produced via structure formation shocks suffer strong collisional losses in the ICM \citep{JubelgasEtAl2008}. Therefore, in general, CRs are not dynamically dominant in galaxy clusters. In particular, the amount of CRs within galaxy clusters has been severely constrained by the non-detection of gamma-ray radiation of clusters to be less than $10-20$ per cent \citep{AckermannEtAl2010, AhnenEtAl2016, BrunettiEtAl2017}. Although CRs are not energetically dominant for clusters as a whole, locally they could still play an important role, e.g., within AGN jets for which the energy composition is largely unknown.  

% Roles of CRs AGN feedback in clusters; observational evidence from bubble compoisiton, early simulations including CRs

The primary motivation for considering the effects of CRs in the context of AGN feedback can be traced back to early observations of cluster radio bubbles. Assuming that the AGN jet-inflated bubbles are in approximate pressure equilibrium with the ambient ICM, it was found that for some of the cluster radio bubbles, the total bubble pressure inferred from the ICM ($\equiv P_{\rm ext}$) is much greater than the internal pressure from radiating CR electrons (CRe) estimated from the radio observations ($\equiv P_{\rm int}$) \citep{MorgantiEtAl1988, HardcastleEtAl1998, DunnFabian2004, RaffertyEtAl2006, BirzanEtAl2008}. Generally speaking, the AGN bubbles could be supported by magnetic fields, thermal gas, CR protons (CRp), and CRe. Magnetic pressure typically is in rough equipartition with the radiating CRe and hence likely subdominant. The low X-ray surface brightness within the AGN bubbles puts stringent constraints on the amount of thermal gas with a temperature of several keV \citep[][]{NulsenEtAl2002}. Therefore, these observations suggest that the radio bubbles with high $P_{\rm ext}/P_{\rm int}$ ratios are likely supported by either ultra-hot thermal gas or non-radiating CRp. While feedback from thermally dominated bubbles inflated by kinetic-energy-dominated jets is more extensively studied (Section~\ref{sec:macro}), the effects of CR-dominated bubbles are less well understood. 

\subsubsection{Impact of Cosmic Rays on AGN Feedback Processes}
\label{sec:cr_feedback}

%In terms of AGN feedback, CRp and thermal gas have several distinct properties. 

In terms of AGN feedback, jet-inflated bubbles filled with thermal gas versus CRs are expected to exhibit distinct properties that could potentially affect the long-term evolution of CC clusters. First, as CRs are relativistic charged particles, they have a softer equation of state (EoS) than thermal gas, i.e., with an adiabatic index of $\gamma_{\rm cr} = 4/3$ (in the ultra-relativistic limit; see \cite{EnsslinEtAl2007} for a more general formalism) rather than $\gamma_{\rm gas}=5/3$ as typically assumed for ideal, monatomic gas. This means that the enthalpy (given by equation \ref{eq:enthalpy}) of CR-dominated lobes would be $4PV$ compared to $2.5PV$ for a non-relativistic fluid \citep{McNamaraNulsen2007}, suggesting that CR dominated lobes would do less work on their environment. The different adiabatic index would alter the compressibility of the fluid when CRs are mixed with the thermal gas. Second, because of the additional pressure support from CRs, AGN bubbles filled with CRs are generally less dense and more buoyant, which may uplift the ICM more efficiently than thermal bubbles. Last but not least, the CRs are expected to provide heat to the ICM via both collisional heating (including Coulomb and hadronic processes) as well as collisionless heating (through the CR streaming instability; more details below). All of the above have motivated the investigation of the effects of CRs in the context of AGN feedback. 

% Brief review about CR physics

Regarding modelling the effects of CRs in fluid simulations, there have also been substantial developments over the past decade. In particular, the CR hydrodynamic/MHD framework has matured and become one of the primary numerical tools for studying the roles of CRs in galaxies and clusters (see review by \citet{Zweibel2013, Zweibel2017}). In this framework, the CRs are treated as a second fluid, and an additional equation for evolving the CR energy density is solved together with the standard fluid equations, with an extra term $\nabla P_{\rm cr}$ in the momentum equation describing the pressure force from CRs. The underlying assumption behind this fluid treatment is that the CRs are well scattered by self-excited waves in the magnetised plasma via the streaming instability \citep{KulsrudPearce1969, Wentzel1974}, i.e., the ``self-confinement picture'' of CR transport, or by waves as part of the cascade of MHD turbulence, i.e., the ``extrinsic turbulence picture.'' In the self-confinement picture, the CRs can stream relative to the gas down their pressure gradients along magnetic fields with the Alfv\'en velocity, $\mathbf{u}_{\rm s}=-{\rm sgn}(\hat{\mathbf{b}}\cdot \nabla P_{\rm cr}) \mathbf{u}_{\rm A}$, where $\mathbf{u}_{\rm A}$ is the Alfv\'en velocity and $\hat{\mathbf{b}}$ is the unit vector of the magnetic field. Therefore, the transport of CRs can be generally prescribed as advection with the gas plus flux terms associated with the streaming and diffusion of CRs. When CRs stream, there is also net energy transfer from the CRs to the thermal gas via Alfv\'en waves, which corresponds to the collisionless CR streaming heating term, $-\mathbf{u}_{\rm A}\cdot \nabla P_{\rm cr}$. In the extrinsic turbulence picture, on the other hand, the CRs are scattered by forward-propagating and backwards-propagating waves. Under the assumption of balanced turbulence, the transport of CRs from these waves cancels out and thus CRs essentially advect with the gas, and streaming heating vanishes. CR diffusion in the simulations represents not only diffusion due to gyroresonant scattering but also CR transport due to unresolved fluctuations in the magnetic field. Regardless of the CR transport model considered, CRs could heat the ICM via collisional processes such as hadronic and Coulomb interactions. In terms of numerical implementations of the CR hydrodynamic equations, it was recognized in earlier simulations that the CR streaming fluxes would cause spurious oscillations near the extrema of CR pressure distributions unless very stringent simulation timesteps are used \citep{SharmaEtAl2010}. More recently, this numerical difficulty has been overcome by introducing a two-moment method for solving an additional equation for CR fluxes \citep{JiangOh2018, ThomasPfrommer2019}. With the above advancement in treating CR physics in large-scale simulations, significant progress has been made in terms of understanding the impacts of CRs on AGN feedback in clusters.  

% Review on more recent works on CR feedback (steady-state models, 3D CR-MHD simulations), CRs can provide enough heating as long as there is transport, whether CR dominated jets can heat the core but not violate gamma-ray constraints

\begin{figure}
\includegraphics[width=13.5cm]{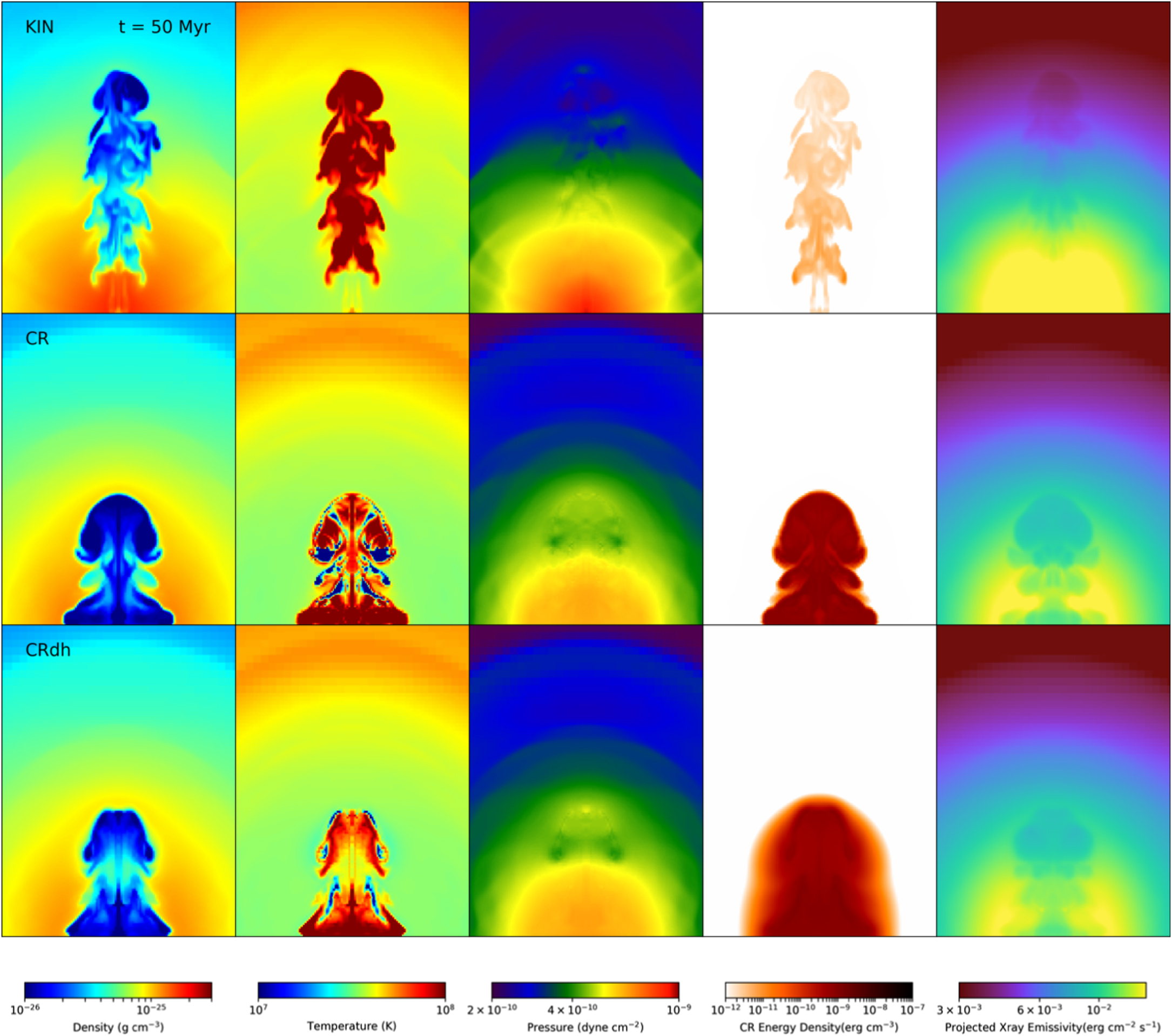}
\caption{Comparisons of simulation results with varied jet composition and assumptions for modeling CR transport. The simulations model a single outburst of jet activity (jet power of $5\times 10^{45}$ erg s$^{-1}$ and jet duration of 10 Myr) in a Perseus-like hydrostatic atmosphere. Rows from top to bottom show the results from kinetic-energy dominated jets (KIN), CR dominated jets (CR), and CR dominated jets with diffusion and heating (CRdh). The morphology of jet-inflated lobes tend to be elongated for the KIN case, whereas bubbles inflated by CR dominated jets are wider. Comparisons between the bottom two panels show that CR heating is more efficient and the amount of cold gas is less for the CRdh case, as the CRs can diffuse outside the bubbles and heat the ICM due to hadronic and Coulomb interactions. (Figure 3 from \citet{YangEtAl2019}, \copyright AAS, reproduced with permission)}
\label{fig:yang_crs}
\end{figure}

CR heating of the ICM has long been considered a possible source to suppress radiative cooling in CC clusters, as the CRs injected by the AGN jets could diffuse or stream outside the bubbles and interact with the ICM via collisional and collisionless processes. While early simulations were not able to demonstrate stabilization of the CCs partially due to relatively simple setups as well as the omission of streaming heating \citep{BoehringerMorfill1988, LoewensteinEtAl1991, RephaeliSilk1995, ColafrancescoEtAl2004, JubelgasEtAl2008, PfrommerEtAl2007}, it was then realised that CR streaming heating via Alfv\'en waves could be a viable mechanism for heating the ICM, and the amount of CR heating could be sufficient to counteract radiative cooling and stabilize the CCs \citep{GuoOh2008, EnsslinEtAl2011, FujitaOhira2011, FujitaOhira2013}. In addition to providing heat to the ICM, the CRs also affect the gas dynamics differently from thermal gas. Comparing AGN bubbles inflated by kinetic-energy dominated jets versus CR-energy dominated jets, the former tends to be more elongated, while CR-dominated bubbles tend to be more oblate due to their lower momenta \citep{GuoOh2008, SijackiEtAl2008, GuoMathews2011, YangEtAl2019}, as shown in Figure \ref{fig:yang_crs}. While AGN bubbles inflated by kinetic-energy dominated jets are easily deformed by fluid instabilities due to the large shear velocity relative to the ambient medium, CR jet-inflated bubbles are generally more stable \citep{SijackiEtAl2008, YangEtAl2019}. The morphology of the CR-filled bubbles can thus more easily explain the ``fat'' bubbles with smooth surfaces seen in observations of X-ray cavities, such as the young bubbles near the centre of the Perseus cluster. The larger cross-sections of CR bubbles, in addition to the buoyancy of the CRs, result in more efficient uplift of the ICM. The outward mass transfer driven by CR bubbles alone could bring the fast-cooling gas away from the core and help reduce cooling near the cluster centre \citep{SijackiEtAl2008, MathewsBrighenti2008, WeinbergerEtAl2017, YangEtAl2019}. It is worth noting that the dynamical effects arising from having CR-energy dominated jets are in many aspects similar to the light jets discussed in Section~\ref{sec:morphology_direction_energetics} because of their lower momenta compared to kinetic-energy dominated, heavy jets. However, because of the differences in the EoS and heating mechanisms between CRs and thermal gas, the detailed heating processes driven by CR jets and light jets may still be different. 

% Steady-state models to 3D dynamic models

While steady-state models have shown that CR heating could be a viable mechanism for stabilizing the CCs \citep{FujitaOhira2013, JacobPfrommer2017a, JacobPfrommer2017b}, recent 3D CR hydrodynamic/MHD simulations have brought rich information about the interplay between the CRs and the ICM. \citet{EhlertEtAl2018} performed 3D MHD simulations of a single AGN outburst with varied jet parameters and demonstrated that CRs could reproduce observed diverse morphologies of radio lobes and provide sufficient heating to the ICM. They identified jet energy as the critical parameter for determining the bubble morphology and CR distribution, while jet luminosity is responsible for setting the Mach numbers of shocks. Using 3D MHD simulations of self-regulated AGN jets dominated by CRp, \citet{RuszkowskiEtAl2017b} showed that the key to successful CR feedback is the CR transport mechanisms, such as CR diffusion and streaming. With these CR transport processes, CRs could then diffuse or stream outside the low-density bubbles and get into contact with the denser ambient ICM, allowing CR heating to operate (shown in the middle and bottom rows of Figure \ref{fig:yang_crs}). Without CR transport (i.e., if only CR advection is considered), the amount of CR heating would be insufficient, and cooling catastrophes would still occur \citep{RuszkowskiEtAl2017b, WangEtAl2020}. It was also found that, due to the significant outward mass transfer driven by CR bubbles, it would take longer for the cluster to contract again and trigger another cycle of AGN activity. As a result, the feeding of the central SMBH and hence the AGN activity tends to be more episodic when the cluster is regulated by CR feedback \citep{RuszkowskiEtAl2017b}, whereas the AGN activity tends to be more quasi-continuous for a similar setup with kinetic-jet feedback \citep{YangReynolds2016Hydro}. This implies that the composition of AGN jets is a major source of uncertainties in simulation predictions for cluster evolution as well as SMBH growth. 

Most of the simulations of CR feedback have focused on CRp-dominated jets; however, observational constraints on the composition of cluster radio bubbles suggest that the distribution is bimodal \citep{DunnFabian2004, RaffertyEtAl2006, CrostonEtAl2018}. For some bubbles, the internal pressure from radiating CRe is much lower than the external pressure of the ICM ($P_{\rm ext}/P_{\rm int} \gg 1$), implying significant pressure support from thermal gas or non-radiating particles. For other bubbles with $P_{\rm ext}/P_{\rm int} \sim 1$, the CRe pressure is enough to keep the bubbles in pressure equilibrium with the surrounding ICM. The latter case has been recently investigated by \citet{LinEtAl2023}. By comparing the feedback effects of CRp and CRe-dominated jets, they found that, although CRe cool much faster than CRp due to IC and synchrotron radiation, the CRe-dominated bubbles quickly become thermally dominated $\sim 20$ Myr after injection. Consequently, the CRe bubbles would not deflate, but they behave similarly to thermal bubbles both in terms of dynamics and ICM heating. This also implies that, for bubbles inflated by CRe jets, there should be a transition of its composition from $P_{\rm ext}/P_{\rm int} \sim 1$ to $P_{\rm ext}/P_{\rm int} \gg 1$. Future measurements of this pressure contrast as a function of radius for cluster radio bubbles may be a viable way to probe the intrinsic composition of AGN jets/bubbles.   

While the above simulations have focused on the properties of the hot ICM, more recent studies have looked into how CR-jet feedback could help suppress SFRs and quench central galaxies. Using the FIRE (Feedback In Realistic Environments) simulations of idealised clusters of $10^{12}-10^{14}\ M_\odot$ with toy models for AGN feedback, \citet{SuEtAl2020} showed that both turbulent stirring and CR heating are efficient mechanisms for maintaining a stable, low-SFR halo over billion-year timescales, in which the core density and cooling rates are suppressed due to the non-thermal pressure support from turbulence or CRs. Later simulations of bi-polar jet feedback by \citet{SuEtAl2021} further performed a thorough parameter study on the relevant jet parameters, including the jet energy composition (kinetic, thermal, CR, or magnetic), jet widths, jet precession, injected mass fluxes, and duty cycles. They showed that CR-dominated jets can most efficiently quench the central galaxy, whereas kinetic jets are less efficient unless they have wide opening angles or precession angles. They attributed the efficient CR feedback to three key factors. First, the injected CRs form wider cocoons that suppress inflows and provide pressure support to the gas, resulting in longer cooling times. Second, as shown in recent simulations of thermal instabilities \citep{JiEtAl2020, ButskyEtAl2020}, the cooling gas would remain diffuse due to the pressure support from CRs, slowing down the ``precipitation'' of cold gas that would be otherwise accreted onto the central black hole. Lastly, because CRs can diffuse and stream through the gas, they tend to generate more gentle outflows rather than explosive ones expected for a pressure-driven blastwave. As a result, CR feedback is less likely to overheat the cluster cores compared to injections in other energy forms.   

\subsubsection{Observational Signatures and Constraints}
\label{sec:cr_obs}

% Observational constraints and signatures

Any successful CR feedback model must satisfy observational constraints of galaxy clusters. The CRs injected from the AGN jets are expected to produce non-thermal emission that may be detected using multi-wavelength observations. For CRe, they could radiate via synchrotron emission when they interact with magnetic fields as well as via inverse-Compton (IC) scattering of ambient photons (e.g., the cosmic microwave background, CMB) --- this is often called ``leptonic emission.'' For CRp, their inelastic collisions with ambient nuclei could produce neutral pions, which decay into gamma rays --- the so-called ``hadronic interaction.'' The hadronic process could also produce charged pions, which decay into charged muons and subsequently to secondary electrons and positrons, together with the production of neutrinos. The secondary particles could again emit due to the synchrotron and IC radiation.

Radio observations near the cluster cores could potentially provide constraints on the amount and distributions of CRs injected by AGN. Indeed, some clusters are known to host radio mini-halos \citep[e.g.,][]{GittiEtAl2007, GiacintucciEtAl2014, RichardLaferriEtAl2020}, which are faint, diffuse radio structures extending $\sim 50-300$ kpc, characterized by steep radio spectra with typical spectral indices of $\alpha > 1$ (as defined in $S_\nu \propto \nu^{-\alpha}$, where $S_\nu$ is the flux density, and $\nu$ is the frequency). The physical origin of the radio mini-halos is not yet fully understood. Generally speaking, because the cooling times of CRe are typically shorter than the time it takes to transport them to the extent of the observed mini-halos, these CRe would need to be produced in situ or be re-accelerated from an old seed population. In terms of the emission mechanisms, the mini-halos could be produced by CRe either via hadronic interactions between CRp and the ambient ICM \citep[e.g.,][]{PfrommerEnsslin2004, FujitaEtAl2007, ZandanelEtAl2014} or via turbulence reacceleration, e.g., by sloshing motions \citep{GiacintucciEtAl2011, ZuHoneEtAl2013} or AGN activity \citep{BraviEtAl2016}). For the latter scenario, \citet{JacobPfrommer2017b} have investigated in detail the predicted radio emission from a steady-state model of CR plus conductive heating. From comparisons with an observed sample of mini-halos, they showed that CC clusters hosting mini-halos cannot be heated by CRs but CCs without mini-halos can. To this end, they proposed a scenario where CC clusters with recent AGN injections should host radio ``micro-halos''; after the CRs diffuse or stream out of the cores, CR heating becomes insufficient and mini-halos light up due to secondary particles produced via hadronic interactions. More simulations on this front are needed in order to verify this picture as well as to understand the source of the seed CRe and the interplay between AGN feedback and sloshing motions in clusters \citep[e.g.,][]{ZuHoneEtAl2021a, ZuHoneEtAl2021b}. Recent radio observations at low frequencies by LOFAR\footnote{\href{https://lofar-surveys.org}{https://lofar-surveys.org}} have also enabled constraints on the interaction between the CRs and the ICM. For example, \citet{BrienzaEtAl2021} found signatures of old AGN jet-inflated bubbles that retain ``mushroom-like'' structures over hundreds of Myr timescales. The bubbles are not thoroughly mixed with the ambient ICM, suggesting suppression of fluid instabilities and CR diffusion at the bubble surface, probably under the influence of magnetic fields.

As mentioned in Section~\ref{sec:cr_motivations}, the non-detection of gamma-ray emission from galaxy clusters has strongly limited the amount of hadronic CRs to be within a few per cent compared to the thermal pressure for clusters as a whole \citep{AckermannEtAl2010, AhnenEtAl2016}. Cosmological simulations by \citet{VazzaEtAl2013} investigated quasar-, jet-, and radio-mode feedback and found successful models that could reproduce X-ray observations of the cluster gas and yield gamma-ray emission below the upper limits by Fermi. Other simulations including CR-dominated jets \citep{RuszkowskiEtAl2017b, YangEtAl2019} showed that, for CR heating to balance radiative cooling, only a small amount of CR pressure support is required. In their successful self-regulated models, CR streaming is a critical ingredient as it would act to remove energy from CRs and heat the gas, reducing the amount of CR pressure support to levels consistent with the gamma-ray constraints. Recent simulations by \citet{BeckmannEtAl2022CR} have also investigated the roles of CR-dominated jets in terms of ICM heating and observable signatures. Consistent with previous works \citep{JiEtAl2020, ButskyEtAl2020}, they found that CRs can modify the development of thermal instabilities and help maintain gas in the hot and warm phases. However, they found that their simulations including CR-dominated jets would produce gamma-ray emission in excess of current observational limits due to the formation of an extended, CR-pressure supported warm nebula, while AGN jets with lower CR fractions ($\sim 10\%$) are allowed and could successfully halt strong cooling flows. The discrepancies in the results have not been fully understood yet, but here we list two possible reasons that might play a role. First, the simulations in \citet{RuszkowskiEtAl2017b} used the mass dropout technique for the cold gas, while \citet{BeckmannEtAl2022CR} retain the cold gas in the simulations. Since the gamma-ray emission produced by hadronic interactions is proportional to the product of the CRp number density and the gas density, it is conceivable that the latter simulation would produce greater gamma-ray emission if the contribution from the cold gas is significant. Another difference between these simulations is that, in contrast to typical AGN feedback prescriptions that connect the jet properties with the black hole accretion rates, \citet{BeckmannEtAl2022CR} additionally modelled the influence of the black hole spin on the jet directions, as discussed in Section~\ref{sec:morphology_direction_energetics}. To this end, their jets are more randomly oriented and could not penetrate and deposit heat to larger radii compared to previous self-regulated simulations with jet precession along a fixed direction. This might have caused less effective feedback and the CR-pressure supported warm nebula that produces the excess gamma-ray emission. More detailed studies are needed in order to pin down this issue. Regardless, the observational gamma-ray limits will provide crucial constraints on the amount of CRs allowed in the cluster cores. 

\subsection{Roles of Plasma Physics of the Intracluster Medium}
\label{sec:plasma}

% Basic properties of the intracluster plasma

While a lot of our understanding of AGN feedback has been gained via fluid simulations, the ICM is in fact a weakly collisional, magnetised plasma. The Coulomb mean free path of the ionized gas in the ICM is \citep{Spitzer1962}
\begin{equation}
\lambda_{\rm mfp} \equiv \frac{3^{3/2}(k_{\rm B}T)^2}{4\pi^{1/2}n_{\rm e}e^4\ln \Lambda} \approx 23\ {\rm kpc} \left( \frac{T}{10^8\ {\rm K}} \right)^2 \left( \frac{n_{\rm e}}{10^{-3}\ {\rm cm}^{-3}} \right)^{-1},
\end{equation}
where $n_{\rm e}$ is the electron number density and $T$ is the temperature of the ICM. Typical values for $\lambda_{\rm mfp}$ range from $\sim 0.1$ kpc near cluster cores to $\sim 10$ kpc in cluster outskirts. The ratio between the collisional mean free path and the size of the system is thus $\lambda_{\rm mfp}/L \sim 0.1-10^{-3}$. For regions in clusters with $\lambda_{\rm mfp} \ll L$, the ICM could be safely approximated as a collisional fluid; for other regions where $\lambda_{\rm mfp} \lesssim L$, the ICM is instead weakly collisional. In addition, as discussed in Section~\ref{sec:macro_environment}, the ICM is magnetised with typical magnetic field strengths on the order of $1-10\ \mu$G \citep{CarilliEtAl2002}. Even though the magnetic field is not dynamically dominant (plasma beta $\beta \equiv P_{\rm th}/P_{\rm B} \sim 100$), it significantly constrains the motions of the charged particles within the ICM as the Larmor radius of their gyro motions ($r_{\rm g}$) is typically more than ten orders of magnitude smaller than the collisional mean free path, i.e., $r_{\rm g} \ll \lambda_{\rm mfp} \lesssim L$. As a result, transport processes such as viscosity and thermal conduction in this weakly collisional, magnetised ICM are expected to be anisotropic along magnetic field lines, whereas the perpendicular transport is significantly suppressed. Also, the parallel transport coefficients along field lines are mediated by plasma physics that occur on the microscopic scales of the particles' gyroradii, i.e., the microphysics. Over the past decade, there has been substantial progress in terms of our understanding of the microphysical plasma processes. These developments and the plasma properties of the ICM have been summarized in a recent review article \citep{KunzEtAl2023}. Here we will focus on the discussion about how these plasma effects could alter the standard hydrodynamic picture of AGN feedback as described in the previous sections.

\subsubsection{Influence of Thermal Conduction on AGN Feedback}
\label{sec:conduction}

Thermal conduction has long been proposed as one of the heating mechanisms in CC clusters, as it could potentially channel heat from the reservoir of thermal energy in cluster outskirts toward the cluster cores \citep[e.g.,][]{ZakamskaNarayan2003, VoigtFabian2004}. While balancing radiative cooling by conductive heating alone would require fine-tuning \citep[e.g.,][]{BregmanDavid1988, ZakamskaNarayan2003}, it may provide partial heating to the CCs and relieve the burden on the AGN. Indeed, some theoretical models have considered conductive heating together with AGN jet heating \citep[e.g.,][]{GuoOh2009, JacobPfrommer2017a}. However, the fact that thermal conduction occurs anisotropically in the magnetised ICM has introduced further complications. In particular, it is found that anisotropic conduction in the conditions of CCs would trigger the heat-flux driven buoyancy instability (HBI; \citep{Quataert2008}), which would act to reorient magnetic field lines in the direction perpendicular to the temperature gradients. In CCs where the temperature gradients are primarily radial, the HBI would then wrap the field lines in the azimuthal direction, which would shut off conductive heat fluxes from cluster outskirts and potentially worsen the cooling-flow problem \citep{ParrishEtAl2009}.   

Fortunately, later simulations have found that the HBI could likely be circumvented by turbulent motions in the ICM \citep{RuszkowskiOh2010}, which could originate from a variety of sources, including g-modes excited by galaxy motions \citep{RuszkowskiEtAl2011} and AGN jet-driven turbulence \citep{YangReynolds2016Conduction}. These works found that the turbulent motions can efficiently randomize the field lines and resume heat fluxes to an effective Spitzer fraction (the suppression factor compared to the full Spitzer value) of $\sim 1/3$, consistent with that for fully tangled magnetic fields. In this case, both conductive heating and AGN heating contribute to counteracting radiative cooling in the CCs. The amount of conductive heating, though, is likely to be subdominant compared to direct AGN heating, shown by both idealised cluster simulations \citep{YangReynolds2016Conduction} as well as cosmological simulations \citep{KannanEtAl2017}. Recently, simulations by \citet{BeckmannEtAl2022Conduction} that include black-hole-spin driven AGN jets found, instead, that AGN jet-driven turbulence is only able to randomize field lines close to the cluster cores, but the HBI could still operate outside $\sim 50$ kpc and isolate the CCs from conductive heating. To this end, they concluded that conductive heating plays a negligible role in regulating radiative cooling in CCs. Again, the difference in the above results could be due to different implementations of the AGN feedback prescriptions (see discussion in Section~\ref{sec:cr_obs}). In addition, the field-line wrapping effect of the HBI would likely be washed out if other sources of ICM turbulence from cosmic accretion or galaxy motions are included in the simulations. All of these simulations including anisotropic conduction have assumed full Spitzer conductivity along the field lines. If the conductive coefficient in the ICM is significantly suppressed due to microphysical plasma processes, as suggested by recent particle-in-cell (PIC) simulations \citep[e.g.,][]{RobergClarkEtAl2016, RobergClarkEtAl2018, KomarovEtAl2018}, then the contribution from conductive heating would be even more inhibited.  

\subsubsection{Influence of Viscosity on AGN Feedback}
\label{sec:viscosity}

While AGN jet-inflated bubbles in purely hydrodynamic simulations are easily deformed and disrupted by Rayleigh-Taylor and Kelvin-Helmholtz instabilities (see discussion in Section~\ref{sec:lobe_icm}), the observed AGN bubbles are generally more regular and do not exhibit clear signs of instabilities. For instance, the young X-ray cavities near the centre of the Perseus cluster have smooth surfaces, and the more evolved ghost cavity in the northwest direction does not break up but shows a flattened morphology \citep{FabianEtAl2005}. These observed properties of the AGN bubbles have posed challenges to purely hydrodynamic modelling of jet-inflated bubbles and called for the consideration of additional mechanisms to suppress the hydrodynamic instabilities such as magnetic fields \citep[e.g.,][]{JonesEtAl2005, Dursi2007, RuszkowskiEtal2007, DursiPfrommer2008, OneillEtAl2009} or viscosity \citep{ReynoldsEtAl2005, SijackiSpringel2006b}. In addition, the morphology of the ${\rm H}_\alpha$ filaments behind the northwest ghost bubble in Perseus suggests that they may be dragged up by the buoyantly rising bubble \citep{FabianEtAl2003}. These observations have motivated the consideration that the ICM may have a non-negligible level of viscosity. 

Indeed, viscous simulations by \citet{ReynoldsEtAl2005} have demonstrated that isotropic viscosity could act to suppress the fluid instabilities and prevent the bubbles from disruption. Additionally, the streamlines behind the simulated bubbles are consistent with the coherent structure of the observed ${\rm H}_\alpha$ filaments. Subsequent work by \citet{SijackiSpringel2006b} also showed that the properties of AGN jet-inflated bubbles, including their morphology, maximum distance from the cluster centre, and survival time, depend sensitively on the level of ICM viscosity. However, this is not the full story because viscosity is expected to be anisotropic in the magnetised ICM, and the parallel viscosity coefficient mediated by microphysical plasma processes should be considered when modelling the evolution of AGN bubbles. To this end, simulations by \citet{DongStone2009} studied the buoyant evolution of initially static bubbles including anisotropic viscosity along magnetic fields of different initial geometries. They found that anisotropic viscosity can efficiently suppress fluid instabilities in the direction parallel to the magnetic field while having little effect on instabilities that develop perpendicular to the field. As a result, the fate of the bubbles is sensitive to the assumed geometry of the magnetic field. While the bubbles can be stabilized by viscosity along initially horizontal fields or toroidal fields confined to the bubble interior, the bubbles are deformed in the case with vertical field geometry. 

\begin{figure}
\includegraphics[width=13.5cm]{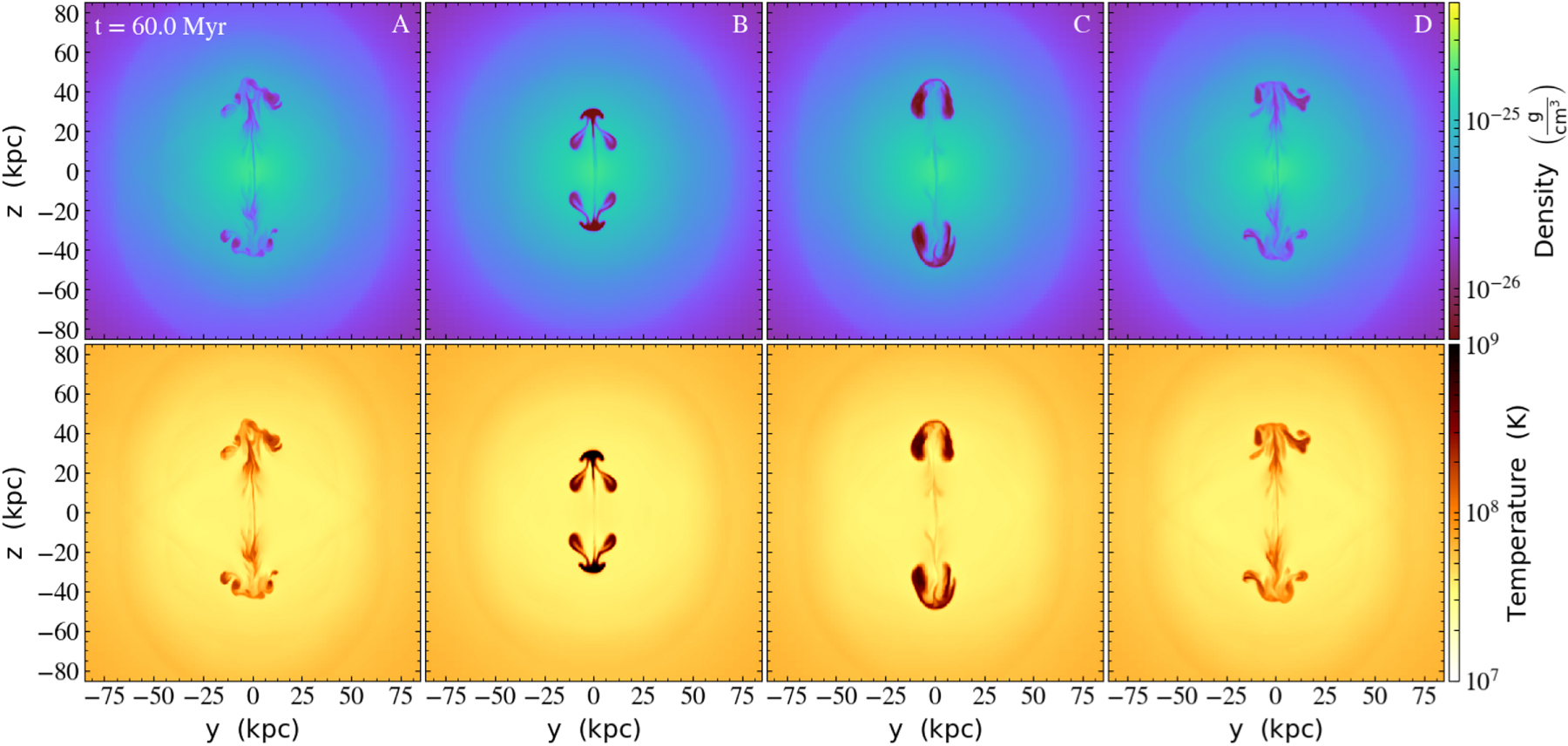}
\caption{Impact of assumptions about ICM viscosity on the evolution of AGN jet-inflated bubbles. Cases A-D show simulations with no viscosity, isotropic viscosity with full Spitzer values, anisotropic viscosity with full Braginskii values, and anisotropic viscosity limited by microinstabilities, respectively. While the hydrodynamic instabilities are suppressed by viscosity in cases B and C, when the parallel viscosity along magnetic field lines is suppressed by microinstabilities (case D), the viscosity is strongly limited and the bubbles are deformed as in the inviscid case (A). This illustrates the importance of modelling the ICM microphysics in AGN simulations. (Figure 1 from \citet{KingslandEtAl2019}, \copyright AAS, reproduced with permission)}
\label{fig:kingsland_viscosity}
\end{figure} 

\citet{KingslandEtAl2019} further investigated the effects of anisotropic viscosity on AGN jet-inflated bubbles in a more realistic tangled magnetic field geometry taking into account microphysical plasma effects learned from recent PIC simulations. Specifically, for a weakly collisional, magnetised plasma like the ICM, the viscosity originates from pressure anisotropy that arises from the conservation of the adiabatic invariants on timescales much greater than the inverse of the ion gyrofrequency \citep{ChewEtAl1956, SchekochihinEtAl2005}. Recent plasma simulations have found that when the pressure anisotropy becomes greater than thresholds that are on the order of $1/\beta$ (where $\beta$ is the plasma beta value), microinstabilities including the firehose and mirror instabilities would be triggered and pin the pressure anisotropy at the marginal-stability thresholds, effectively suppressing the parallel viscosity coefficients \citep[see][and references therein]{KunzEtAl2023}. Incorporating these latest findings into Braginskii-MHD simulations, \citet{KingslandEtAl2019} found that, for anisotropic viscosity with full Braginskii values \citep{Braginskii1965}, the integrity of the bubbles could be preserved because viscosity along tangled field lines could suppress the instabilities in multiple orientations on the bubble surface (see case C in Figure \ref{fig:kingsland_viscosity}). However, when suppression of the parallel viscosity coefficients by the microinstabilities is considered, the suppression is so strong that the bubbles are deformed just as in the inviscid case (case D in Figure \ref{fig:kingsland_viscosity}). This is because the plasma beta value in the bubble interior could be as high as $10^4$, which dramatically limits the amount of pressure anisotropy and viscosity in the vicinity of the bubbles. Therefore, they concluded that Braginskii/anisotropic viscosity is unlikely to be the primary mechanism for preserving the coherence of AGN bubbles, but other mechanisms (e.g., magnetic fields) are still required to reproduce the morphology of observed bubbles. This further emphasized the importance of modelling the ICM plasma properties as the bubble evolution has a direct influence on their ability to uplift the ICM and where the bubbles deposit heat provided by the AGN.

Finally, since the pressure anisotropy in the ICM provides an anomalous ``effective'' viscosity \citep{KunzEtAl2023}, there could be heating associated with parallel viscous dissipation of gas motions (so-called gyroviscous heating). By assuming the pressure anisotropy is pinned at the marginal-stability thresholds in the turbulent ICM, \citet{KunzEtAl2011} showed that this mechanism could provide heating rates comparable to the radiative cooling rates. Recent PIC simulations have further studied the detailed process of how particles can be gyroviscously heated by large-scale turbulent fluctuations via magnetic pumping \citep{LeyEtAl2023}. More studies would be required in order to integrate these microphysical phenomena into large-scale turbulence models of the ICM for a full assessment of gyroviscous heating of the ICM.        

\subsubsection{Heating by Sound-Wave Dissipation}
\label{sec:soundwave}

In X-ray images of the Perseus cluster, there are ripple-like structures that were interpreted as sound waves driven by central episodic AGN outbursts \citep{FabianEtAl2003, SandersFabian2007}. Motivated by this observational finding, early simulations have shown that viscous dissipation of sound waves could be a viable mechanism for heating the ICM \citep{RuszkowskiEtAl2004a, RuszkowskiEtAl2004b}, although some studies found that their energy content may be small due to viscous damping \citep{SijackiSpringel2006b}. Sound-wave heating is an attractive solution to the cooling-flow problem because it could heat the ICM isotropically as the waves propagate from the cluster centre outward. The question of whether sound-wave dissipation could be a significant contributor to ICM heating could be broken down into two: how much of the injected energy by the AGN can be stored in the form of sound waves, and how the waves propagate and dissipate in the ICM. We will summarize the progress along these lines below. 

\begin{figure}
\includegraphics[width=13.5cm]{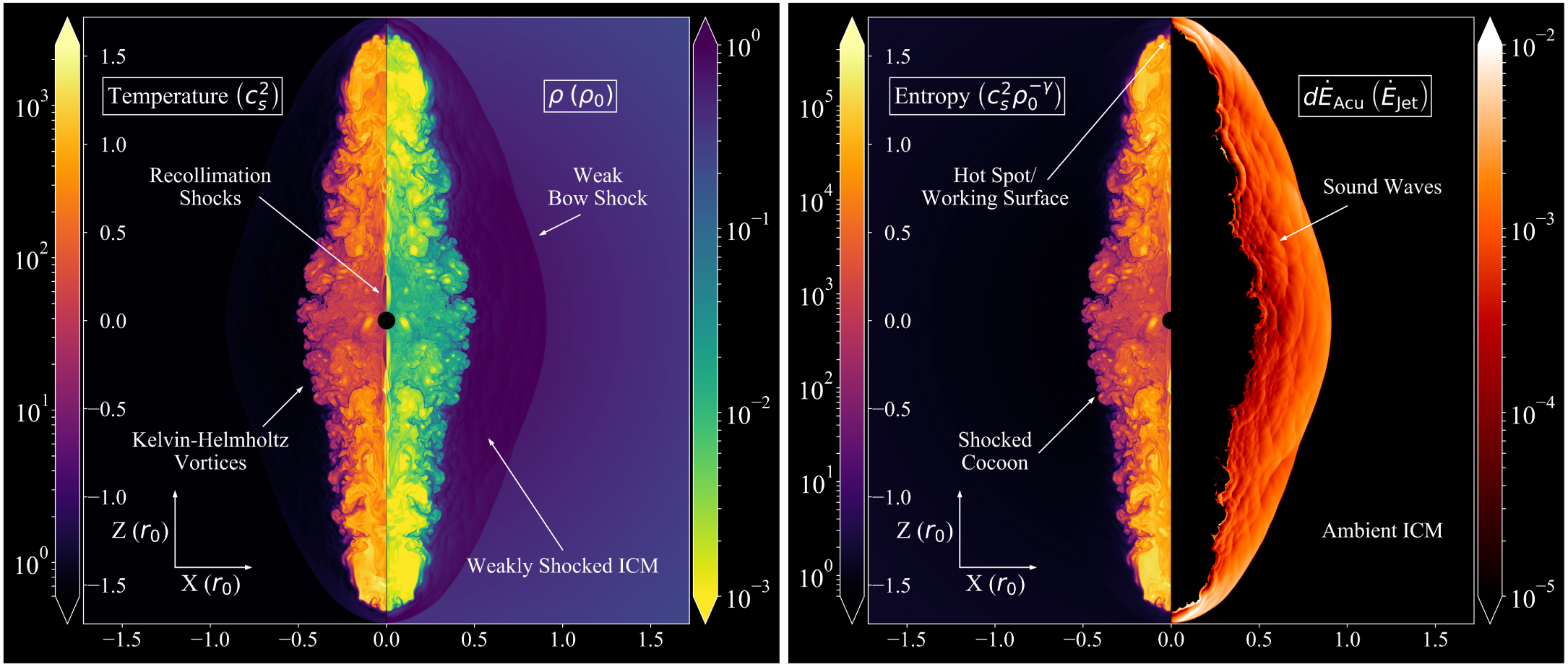}
\caption{Illustration of sound waves generated by jets. The left-hand panel shows the jet and cocoon temperature and density structure, with key features labelled. The right-hand panel shows jet entropy and the acoustic flux density, with structure in the latter illustrating the production of sound waves within shocked ICM material. (Figure 1 from \citet{BambicReynolds2019}, \copyright AAS, reproduced with permission)}
\label{fig:bambic_soundwaves}
\end{figure} 

Several recent simulations have been conducted in order to quantify how much of the injected AGN energy can be converted into sound waves. Using simulations of spherical injections into a uniform medium with varied injected energy and duration, \citet{TangChurazov2017} found that the fraction of AGN energy that goes into sound waves is dependent on the duration of the injection. When the duration is long, in which the solution approaches the ``slow piston'' limit, the energy of the sound waves is close to zero. For instantaneous outbursts (duration approaches zero), the energy fraction of sound waves is $\lesssim 12\%$. \citet{BambicReynolds2019} further performed a parameter study using axisymmetric simulations of an AGN outburst with bipolar jet geometries. Figure~\ref{fig:bambic_soundwaves} is taken from their paper and illustrates the generation of sound waves by a jet. They found that up to $\sim 25\%$ of the injected energy could be stored in the form of sound waves for optimal parameter combinations. Using jet parameters informed by previous self-regulated AGN feedback simulations, \citet{WangYang2022} found that the production efficiency of compressional waves (including weak shocks and sound waves) is $\sim 9\%$ for a single AGN energy injection. However, this production efficiency drops to less than $\sim 3\%$ in self-regulated feedback simulations. This is because, with repeated AGN outbursts, shocks are almost continuously generated and their energy dominates that of the weak shocks. To this end, shock dissipation together with destructive interference of sound waves would act to reduce the amount of energy in compressional waves. Considering the above effects, they concluded that sound-wave heating may be a subdominant source of heating in CC clusters. However, \citet{WangYang2022} have assumed powerful AGN injections via bipolar jets. Using viscous simulations of a single spherical, gentle (longer-duration) AGN outburst, \citet{ChoudhuryReynolds2022} found that $\sim 20\%$ of the injected power can be carried away by sound waves. Overall, the above results suggest that the production efficiency of sound waves is $\sim 3-25\%$, depending on how the feedback energy is injected (e.g., geometry, duration) and how the waves generated by repeated events interfere with each other. Finally, the above estimates are derived based on hydrodynamic treatments of the ICM, and there may be additional plasma effects that need to be taken into account in terms of sound-wave generation. For instance, \citet{KempskiEtAl2020} have performed Braginskii-MHD simulations including CRs, and showed that the phase shifts between CR pressure and density fluctuations would generate instabilities when the CR fraction is greater than a threshold proportional to the plasma beta. Therefore in an ICM with high plasma beta values, this mechanism could potentially be important for exciting sound waves.

Another question to address is the propagation and dissipation of sound waves in the ICM. Pioneering work by \citet{FabianEtAl2005} showed that nearly adiabatic acoustic waves would damp within one wavelength of their source and overheat the cluster core if thermal conductivity and viscosity are at their full Braginskii values. If this were true, it would pose challenges to the interpretation of observed ripples as propagating sound waves. Assuming thermal conduction is completely suppressed and viscosity is reduced to $10\%$ of the Braginskii value, heating due to sound-wave dissipation could then balance radiative cooling for a power-law spectrum of waves. More recently, \citet{ZweibelEtAl2018} revisited this topic by additionally taking into account the self-limiting nature of electron thermal conduction, differences between electron and ion temperatures, and plasma effects from kinetic theories. Though the kinetic effects somewhat suppress damping and mitigate the problem of wave propagation, the conclusions are reinforced that, in order for acoustic waves to propagate to large radii and heat the CC, drastic suppression of the transport coefficients is needed. Such reduced transport could come from increased effective collisionality of the ICM due to magnetic field fluctuations on small scales driven by plasma instabilities \citep[e.g.,][]{RobergClarkEtAl2016, KunzEtAl2014}. However, future works are still required to fully assess the generation and damping of acoustic waves under the influence of plasma instabilities as well as realistic sources and geometries.   

\subsection{Open Questions and Future Opportunities}
\label{sec:micro_future}

% Observational constraints on bubble composition (SZ, IC in X-ray, P_ext/P_int ratios vs. radius?)

Following the discussion in Section~\ref{sec:cr}, it is obvious that constraining the composition of AGN jets and bubbles is one of the key questions to address since different jet compositions would have different dynamical and thermal impacts on the ICM, which would then affect the long term evolution of clusters as well as SMBH accretion histories. One of the proposed ways to distinguish AGN bubbles dominated by ultrahot thermal gas and CRs is by using the SZ effect \citep{SZ1972}, which arises due to the IC scattering of the CMB photons by electrons in the ICM. It was shown by \citet{PfrommerEtAl2005} that, because the SZ increment/decrement at observed frequencies below $\sim 400$ GHz is greater for thermal gas than a power-law distribution of relativistic CRe, the CR-dominated bubbles would show a clear deficit in the SZ signal, creating ``SZ cavities'' similar to the X-ray cavities. In contrast, there should be no obvious SZ cavities associated with thermally dominated bubbles. The level of the SZ deficits is estimated to be $\sim 6\%-9\%$ for sightlines passing through the CR bubbles \citep{YangEtAl2019}. Recent simulations by \citet{EhlertEtAl2019} have also examined potential systematic effects when modelling the SZ signal of AGN bubbles. They found that the cut-out method for identifying bubbles would fail to account for the shock-enhanced pressure cocoon outside the bubbles, especially for small jet inclinations along the line of sight. Also, in this case, the kinetic SZ effects become relevant and need to be modelled. Observationally, the detection of SZ cavities in the cluster MS 0735.6+7421 has been first reported by \citet{AbdullaEtAl2019} and later confirmed by \citet{OrlowskiSchererEtAl2022}, which suggests that the AGN bubbles in MS 0735.6+7421 are supported by the non-thermal pressure provided by CRs, though thermal gas with a temperature greater than $\sim 150$ keV has not yet been ruled out. High-resolution, high-sensitivity SZ observations such as ALMA\footnote{\href{http://almaobservatory.org/en/home/}{http://almaobservatory.org/en/home/}}\citep{ALMAWP2009}, MUSTANG-2\footnote{\href{https://greenbankobservatory.org/science/gbt-observers/mustang-2/}{https://greenbankobservatory.org/science/gbt-observers/mustang-2/}}\citep{Mustang2WP2014}, and NIKA2\footnote{\href{https://ipag.osug.fr/~ponthien/NIKA2/Welcome.html}{https://ipag.osug.fr/~ponthien/NIKA2/Welcome.html}}\citep{Nika22018}, will be able to obtain important constraints on the composition of a larger sample of AGN bubbles in the near future.      

% Observational constraints on the transport coefficients 

Much detail of the AGN feedback processes relies on our understanding of the plasma properties of the ICM. While recent PIC simulations have allowed tremendous progress on the theoretical front, observational constraints are demanded in order to pin down what is the best description for the weakly collisional, magnetised ICM. In particular, constraints on the transport coefficients would be key to understanding ICM properties. For instance, previous X-ray observations of the sharpness of cold fronts have ruled out ICM conductivity at the full Braginskii value \citep[e.g.,][]{ZuHoneEtAl2013b}. Comparisons between simulation predictions and observations also suggest that either full Braginskii viscosity or isotropic Spitzer viscosity with a suppression factor $f_{\rm sp}\sim 10\%$ is consistent with the observed cold fronts \citep{ZuHoneEtAl2015}. Observations of ram-pressure stripping tails of galaxies infalling into the ICM could also be used to probe the viscosity of the ICM, and previous studies \citep[][]{SuEtAl2017, WangMarkevitch2018} have placed constraints on the ICM viscosity at a level of $f_{\rm sp} \sim 5-20\%$ compared to the full Spitzer value (assuming isotropic viscosity). Observations of the ICM turbulence spectrum down to the scale where transport processes become relevant could also yield constraints on ICM viscosity. Using Chandra observations of the Coma cluster and inferring the turbulence spectrum from X-ray surface brightness fluctuations, \citet{ZhuravlevaEtAl2019} found that the ICM viscosity must be strongly suppressed, with an effective isotropic viscosity of $f_{\rm sp} \sim 0.1-10\%$. Improved measurements from upcoming and future X-ray missions including XRISM$^{\ref{note:xrism}}$, AXIS$^{\ref{note:axis}}$, LEM\footnote{\href{https://www.lem-observatory.org}{https://www.lem-observatory.org}}\citep{LEMWP2022}, Athena$^{\ref{note:athena}}$, and Lynx\footnote{\href{https://www.lynxobservatory.com}{https://www.lynxobservatory.com}}\citep{Lynx2019}, will provide crucial constraints on the transport processes in the ICM and hence further understanding about the AGN feedback mechanisms.     

\section{Concluding Remarks}
\label{sec:conclusions}

Thanks to cutting-edge numerical simulations with increasing resolution and complexity in terms of input physics, tremendous progress has been made over the past decade regarding the understanding of the macro- and microphysics of radio jet feedback in galaxy clusters. As reviewed in this article, while extensive studies have reinforced the importance of jet feedback in suppressing cooling flows in massive clusters, many details are not yet fully understood. In particular, the relative importance among various heating mechanisms, including mixing, shocks, sound waves, turbulence, and CRs, as well as their subsequent impact on SMBH feeding and cluster evolution, critically depend on the properties of the jet injections (e.g., composition, duration, direction) and how the lobes are inflated and evolve with time (e.g., whether the hydrodynamic instabilities can be suppressed, whether the bubbles could be preserved to large radii by magnetic fields or viscosity). Some of the open questions and future opportunities have been discussed in Section~\ref{sec:macro_future} and Section~\ref{sec:micro_future}. Here we conclude by highlighting some of the key unresolved questions:

\begin{itemize}
\item{What are the next steps for improving sub-grid AGN models in numerical simulations? Do we require other feedback channels in addition to radio jets to regulate galaxy clusters?}
\item{How exactly is the SMBH feeding and feedback cycle established across such a huge dynamical range?  What is the best way to model the evolution of black hole spins and couple them to jet feedback?}
\item{How does environment (e.g., cluster weather, magnetic fields) impact the effectiveness and mechanisms through which jet feedback couples to the ICM and how does it impact lobes distributions, lifetimes and morphologies?}
\item{What is the composition of AGN jets and bubbles? Are the jets light or heavy? Are they energetically dominated by ultra-hot thermal gas, CRp, CRe, or magnetic field? How does the composition vary (e.g., with morphological types, environment, launching mechanisms)? How does the jet composition impact the AGN feeding and feedback processes?}
\item{How important are CRs in the process of AGN feedback and how can it be constrained using observations of the non-thermal emission they produce? What is the best way to model CR transport in the turbulent, magnetised ICM?}
\item{What is the valid prescription for modelling the weakly collisional, magnetised ICM plasma? What are the levels of thermal conductivity and viscosity in the ICM?}
\item{Can we extrapolate our knowledge about cluster feedback down to scales of galaxy groups and elliptical galaxies? Does AGN feedback operate in the same way across different mass scales?}\\
\end{itemize}

\authorcontributions{Both authors contributed to Section~\ref{sec:overview} and Section~\ref{sec:conclusions}. M.A.B. was the main contributor to Section~\ref{sec:macro} and H.-Y.K.Y. was the main contributor to Section~\ref{sec:micro}. All authors have read and agreed to the published version of the manuscript.}

\funding{M.A.B. was supported by the Science and Technology Facilities Council (STFC). H.-Y.K.Y. was supported by the National Science and Technology Council (NSTC) of Taiwan (109-2112-M-007-037-MY3) and Yushan Scholar Program of the Ministry of Education (MoE) of Taiwan.}

\acknowledgments{We are grateful to the anonymous reviewers for their constructive reports that helped to improve the clarity and breadth of this work. We also thank Filip Hu{\v{s}}ko, Yuan Li and Chris Bambic for allowing use of figures, as well as Chris Reynolds, Debora Sijacki, Franco Vazza and Sophie Koudmani for useful advice, comments and feedback on the article.}

%%%%%%%%%%%%%%%%%%%%%%%%%%%%%%%%%%%%%%%%%%
\begin{adjustwidth}{-\extralength}{0cm}

\reftitle{References}

\bibliography{refs}

\end{adjustwidth}
\end{document}